\documentclass[12pt, aps, prd, oneside,abstract]{revtex4-2}   	
\usepackage{mwe,float}
\usepackage{caption,subcaption, url}
\usepackage{natbib}
\usepackage[nointegrals]{wasysym}
\usepackage{rotating}
\usepackage[pdftex]{color}
\usepackage{graphicx,bm}
 \usepackage{floatpag} 
 \rotfloatpagestyle{empty}
\usepackage{amsmath}
\usepackage{amsthm}
\usepackage{amssymb}
\usepackage{latexsym}
\usepackage{hyperref,rotating}
\usepackage{xpatch,bigints}
\usepackage{booktabs,subcaption}
\usepackage[pdftex]{color}
\usepackage[dvipsnames]{xcolor}
\makeatletter
\AtBeginDocument{\xpatchcmd{\@thm}{\thm@headpunct{.}}{\thm@headpunct{\enskip}}{}{}}
\makeatother 
\usepackage{gensymb}
\usepackage{mathdots}

\newcommand{\qand}{\quad \mbox{and} \quad}

\newcommand{\bos}{\boldsymbol}

\usepackage[version=4]{mhchem}

\usepackage{graphicx,kecpm,slashed}
\usepackage{makeidx}
\makeindex

\DeclareMathOperator{\arctanh}{arctanh}

\definecolor{navyblue}{rgb}{.05,0,.55}

\newcommand{\grad}{\nabla}

\newcommand{\bex}{\begin{example}}
\newcommand{\eex}{\end{example}}
\newcommand{\besp}{\begin{split}}
\newcommand{\ensp}{\end{split}}

\newcommand{\thet}{\theta}

\newcommand{\La}{\Lambda}

\newcommand{\by}{\times}

\newcommand{\ovl}{\overline}

\newcommand{\tit}{\textit}

\newcommand{\btab}{\begin{tabular}}
\newcommand{\etab}{\end{tabular}}
\newcommand{\barr}{\begin{array}}
\newcommand{\earr}{\end{array}}
\newcommand{\bpm}{\begin{pmatrix}}
\newcommand{\epm}{\end{pmatrix}}
\newcommand{\bit}{\begin{itemize}}
\newcommand{\eit}{\end{itemize}}
\newcommand{\ben}{\begin{enumerate}}
\newcommand{\een}{\end{enumerate}}
\newcommand{\bct}{\begin{center}}
\newcommand{\ect}{\end{center}}

\newcommand{\ra}{\rangle}
\newcommand{\la}{\langle}
\newcommand{\bes}{\begin{split}}
\newcommand{\ens}{\end{split}}

\newcommand{\lt}{\left}
\newcommand{\rt}{\right}

\newcommand{\transpose}
{{\mbox{${}^{\scriptsize{\textsf{T}}}$}}}

\begin{document}
\title{Spinors of Spin-one-half Fields} 
\author{Kevin Cahill}
\affiliation{Department of Physics and Astronomy\\
University of New Mexico\\
Albuquerque, New Mexico 87131}
\date{\today}

\begin{abstract}
This paper reviews how a two-state, spin-one-half system transforms under rotations.  It then uses that knowledge
to explain how momentum-zero,
spin-one-half annihilation and 
creation operators transform under rotations. 
The paper then explains how a spin-one-half
field transforms under rotations. 
The momentum-zero spinors are found
from the way spin-one-half systems
transform under rotations and from 
the Dirac equation.  Once the momentum-zero
spinors are known, the Dirac equation
immediately yields the
spinors at finite momentum.
The paper then shows that with these spinors, 
a Dirac field transforms appropriately 
under charge conjugation, parity, and time reversal. 
\par
The paper also describes how a Dirac field
may be decomposed either
into two 4-component Majorana fields 
or into
a 2-component left-handed field
and a 2-component right-handed field.
Wigner rotations and Weinberg's 
derivation of the properties of spinors
are also discussed.
\end{abstract}
\maketitle

\section{Introduction
\label{Introduction sec}}

A four-component, spin-one-half field
invented by Dirac describes the
quarks and leptons of 
the standard model.
Dirac fields are therefore
of enormous importance
in particle physics as well as in
nuclear and atomic physics,
and in cosmology.
Nearly a century has passed
since Dirac's 
description of spin-one-half fields.
One therefore might expect 
that they would be
explained 
clearly and thoroughly
in all modern textbooks
on quantum field theory.
\par
Not quite.
All modern
textbooks on quantum field theory 
describe to some extent
the annihilation operator
$a(\bos p, s)$, which deletes
a particle of momentum $\bos p$
and spin $s$, and the creation operator
$a_c^\dag(\bos p, s)$, which adds
an antiparticle of momentum $\bos p$
and spin $s$.
They all get the $2\pi$'s and
the phase factors 
$e^{\pm i  p \cdot x}$ right
and provide for the
Dirac field a formula like
\begin{equation}
\psi_{\scriptscriptstyle D}(x) 
={}
\sum_{s=\pm 1/2}
\int \frac{d^3p}{(2 \pi)^{3/2}} 
\Big[
u_{\scriptscriptstyle D}(\bos p, s) \, e^{ip\cdot x} \, a(\bos p,s)
+
v_{\scriptscriptstyle D}(\bos p, s) \, e^{-ip\cdot x} 
\, a_c^\dag(\bos p, s) \Big]
\label {4 component Dirac field I}
\end{equation}
in which $p \cdot x ={} \bos p \cdot \bos x
- p^0 t$, $p^0 = \sqrt{\bos p^2 + m^2}$,
$\hbar = c = 1$, and 
Dirac's index $ {\scriptstyle D}$
runs from 1 to 4.  
But three of the leading textbooks
on quantum field theory~\citep{Ryder:1985wq,Kaku:1993ym,Schwartz:p189}
give incorrect formulas for the spinors
$v_{\scriptscriptstyle D}(\bos p, s) $
that multiply creation operators.
And only five~\citep{Srednicki2007spinors,peskin1995AnIntroQFTspinors,Chen:2018cts,Mandl:1985bg,WeinbergIp=0spinors}
of the 15 leading 
textbooks
give explicit, correct formulas
for the spinors.
\par
The only book that fully
explains what spinors are
and that derives formulas for them
is 
{\textit{The Quantum Theory 
of Fields I}} by Steven
Weinberg~\citep{WeinbergIp=0spinors}\@.
His treatment of this and other topics
is so deep and so general, however,
that he had to skip many
intermediate steps to avoid
having his book run to
several thousand pages.
The paper~\citep{Cahill:2005zb}
by Peter Cahill and me
fills in some of these steps.
\par
My purpose in this paper
is to point out that 
one may use the Dirac equation
and elementary
quantum mechanics, specifically
how the states of a spin-one-half
system of momentum zero
transform under rotations, 
to derive 
explicit formulas 
for the spinors
$u_{\scriptscriptstyle D}(\bos p, s)$
and 
$v_{\scriptscriptstyle D}(\bos p, s) $.
\par
The Dirac equation in momentum space
yields the spinors at finite momentum
$u_{\scriptscriptstyle D}(\bos p, s)$
and 
$v_{\scriptscriptstyle D}(\bos p, s) $
once we know the spinors at 
momentum zero
$u_{\scriptscriptstyle D}(\bos 0, s)$
and 
$v_{\scriptscriptstyle D}(\bos 0, s) $, 
but it does not
tell us what the spinors are
at momentum zero.  It merely
tells us that the spinors
$u_{\scriptscriptstyle D}(\bos 0, s)$
are eigenstates of $\c^0$
with eigenvalue $-i$,
and that the spinors 
$v_{\scriptscriptstyle D}(\bos 0, s)$
are eigenstates of $\c^0$
with eigenvalue $i$\@.
But the eigenvalues $-i$ and $i$ 
are degenerate; each has two
eigenvectors.
It is the way a nonrelativistic
spin-one-half system
transforms under rotations
that tells us both which of
the degenerate eigenvectors
of $\c^0$ with eigenvalue $-i$
is $u_{\scriptscriptstyle D}(\bos 0, \thalf)$
and which is
$u_{\scriptscriptstyle D}(\bos 0, -\thalf)$,
and also
which of
the degenerate eigenvectors
of $\c^0$ with eigenvalue $i$
is $v_{\scriptscriptstyle D}(\bos 0, \thalf)$
and which is
$v_{\scriptscriptstyle D}(\bos 0, -\thalf)$.
The Dirac equation then immediately
gives us the spinors at
finite momentum
$u_{\scriptscriptstyle D}(\bos p, s)$
and 
$v_{\scriptscriptstyle D}(\bos p, s) $.  
This derivation is the
simplest one I know of and the
one that most reflects the
way spin-one-half systems
behave under rotations.
\par
Some of the notation used in this paper
and elsewhere
is described in Section~\ref{Notation sec}.
Section~\ref{Rotations of states of zero momentum and spin one-half sec} reviews how 
the Pauli matrices represent 
rotations of nonrelativistic,
spin-one-half systems. 
This knowledge is applied in
Section~\ref{Rotations of spin-one-half creation and annihilation operators of momentum zero sec} 
to two nonrelativistic, spin-one-half
systems:
the momentum-zero annihilation operators 
$a(\bos 0, s)$ for $s = \pm \thalf$
and the momentum-zero antiparticle
creation operators $a^\dag_c(\bos 0, s)$\@.
This understanding of how
spin-one-half creation and annihilation operators of momentum zero transform under rotations is used in
Section~\ref{Rotations of two-component spin-one-half fields sec} to explain how rotations
about the $z$ axis transform
the two 2-component fields
that make up a 4-component Dirac field 
and to determine 
their momentum-zero, 2-component 
spinors $u_\a(\bos 0, s)$ and $v_\a(\bos 0, s)$
for $\a = 1, 2$\@.
The Dirac equation in momentum space
at momentum zero
is used in
Section~\ref{Four-component spinors at zero momentum sec} 
to determine how the four 
momentum-zero 2-component 
spinors $u_\a(\bos 0, s)$ and $v_\a(\bos 0, s)$
are combined into 
two momentum-zero,
4-component Dirac spinors
$u_{\scriptscriptstyle D}(\bos 0, s)$
and 
$v_{\scriptscriptstyle D}(\bos 0, s) $\@.
Section~\ref{Four-component spinors at finite momentum sec}
shows how to use the Dirac equation 
in momentum space to determine
the 4-component Dirac spinors
$u_{\scriptscriptstyle D}(\bos p, s)$
and 
$v_{\scriptscriptstyle D}(\bos p, s) $
at momentum $\bos p$
from the spinors
$u_{\scriptscriptstyle D}(\bos 0, s)$
and 
$v_{\scriptscriptstyle D}(\bos 0, s) $
at momentum zero.
These Sections
are aimed at students who have had a
good  course in quantum mechanics 
at the level of the book 
\tit{Modern Quantum Mechanics} by
Sakurai~\citep{Sakurai1994}\@.
\par
The two 2-component fields
that make up a 4-component Dirac field 
transform the same way under rotations
but differently
under Lorentz boosts.
Section~\ref{Left-handed and right-handed spin-one-half fields sec}
explains why these 2-component
fields are called left handed
and right handed and why 
at high momentum
left-handed fields 
annihilate particles
whose momenta are
antiparallel to their spins
and create antiparticles
whose momenta are 
parallel to their spins
while 
right-handed fields 
annihilate particles
whose momenta are
parallel to their spins
and create antiparticles
whose momenta are 
antiparallel to their spins.
This Section and the three appendices
that follow it are at the graduate level. 
\par
Appendix~\ref{Charge conjugation, parity, and time reversal sec} uses the properties
of spinors as explained in Sections
~\ref{Four-component spinors at zero momentum sec} and
\ref{Four-component spinors at finite momentum sec}
to derive the
behavior of a Dirac
field under parity,
charge conjugation,
and time reversal. 
Appendix~\ref{Wigner rotations sec} 
shows that the requirement
that a Dirac field transform
properly under Lorentz 
transformations determines 
the spinors.
Appendix~\ref{Majorana and Dirac Fields sec} describes Majorana 
and 2-component fields.

\section{Notation}
\label{Notation sec}

This Section describes some  
symbols used in this paper
and others that are merely worth knowing.

\begin{description}

\item[Spin one-half] 
When measured along any axis, 
the spin of a spin-one-half particle
is $\pm \hbar/2$.   

\item[Units] 
$\hbar = c = 1$.

\item[Kronecker delta]
The Kronecker delta is
\begin{equation}
\d_{s,s'} = \lt \{ 
\begin{array}{ll}
1 & \quad \text{if} \>\; s = s' \\
0 & \quad \text{if} \>\; s \ne s' 
\end{array}
\rt. .
\end{equation}

\item[Metric]
The metric of flat spacetime is
the $4\by4$ diagonal matrix $\eta$
\begin{equation}
\eta ={} 
\begin{pmatrix}
-1 & 0 & 0 & 0 \\
0 & 1 & 0 & 0 \\
0 & 0 & 1 & 0 \\
0 & 0 & 0 & 1 
\end{pmatrix} .
\end{equation}
Its elements are 
$\eta_{00} ={} \eta^{00} = -1$,
and 
$\eta_{ik} = \eta^{ik} = \d_{ik}$
for $i, k = 1, 2, 3$
with $\eta_{i0} = \eta^{i0}
= \eta_{0i} = \eta^{0i} = 0$.
Many authors use ${} - \eta$
instead.

\item[4-vectors]
The 4-vectors of position 
and momentum are
$x = (x^0, x^1, x^2, x^3)
= (t, \bos x)$ and
$p = (p^0, p^1, p^2, p^3)
= (E, \bos p)$.
With lowered indexes, they are
$(x_0, x_1, x_2, x_3) = (-t, \bos x)$
and
$ (p_0, p_1, p_2, p_3) = (-E, \bos p)$.

\item[Summation convention]
A repeated index often is meant to
be summed over, as in
\begin{equation}
p \cdot x = {}
p^i x_i = p_i x^i 
= \sum_{i=0}^3 p_i \, x^i
= \bos p \cdot \bos x - p^0 x^0
\qand
x_i = \eta_{ik} \, x^k.
\end{equation}

\item[Dirac delta function]
The functional $\d(x-y)$
maps the function $f(x)$
to the number $f(y)$
\begin{equation}
f(y) = \int dx f(x) \, \d(x-y) .
\end{equation}

\item[Dirac notation]
A state that represents
a particle of kind $n$,
momentum $\bos p$,
and spin $s$ in the $z$ direction
called a ket and is written
as $ | \bos p, s, n \ra$.
The hermitian adjoint
of the ket $ | \bos p', s', n' \ra $ is the
bra $\la \bos p', s', n'  | $.
Their inner product
is $ \la \bos p', s', n'  | \bos p, s, n \ra 
= \d^{(3)}( \bos p - \bos p ') 
\d_{s,s'} \d_{n n'}$.

\item[$d^3p$]
The differential $d^3p$ is $dp_x dp_y dp_z$
or equivalently $dp_1 dp_2 dp_3$.

\item[Annihilation operator]
The annihilation operators 
$a(\bos p, s)$ and $a_c(\bos p, s)$ 
respectively delete
from a state either
a particle or an antiparticle
of momentum $\bos p$
and spin $s$ in the $z$ direction.
The energy of the particle or 
antiparticle is
$p^0 = \sqrt{\bos p^2 + m^2}$
where $m > 0$ is the mass of the particle.
All the 
annihilation operators of a field
delete particles of the same mass.

\item[Creation operator]
The creation operators 
$a^\dag(\bos p, s)$ and
$a^\dag_c(\bos p, s)$ 
respectively add
to a state either
a particle or an antiparticle
of momentum $\bos p$
and spin $s$ in the $z$ direction.
The energy of the particle or
antiparticle is 
$p^0 = \sqrt{\bos p^2 + m^2}$
where $m > 0$ is the mass of the particle.
All the 
creation operators of a field
add particles of the same mass.

\item[Spinors]
The spinor
$u_{\scriptscriptstyle D}(\bos p, s)$
has four components, 
$ {\scriptstyle D} ={} 1, 2, 3, 4$.
It is the coefficient
of the annihilation operator 
$a(\bos p, s)$ in the
Fourier expansion 
(\ref{4 component Dirac field I})
of a Dirac field.
The spinor
$v_{\scriptscriptstyle D}(\bos p, s)$
has four components, 
$ {\scriptstyle D} ={} 1, 2, 3, 4$.
It is the coefficient
of the creation operator 
$a^\dag(\bos p, s)$ in the
Fourier expansion 
(\ref{4 component Dirac field I})
of a Dirac field.

\item[Dirac field]
The Dirac field 
(\ref{4 component Dirac field I})
has four components
${\scriptstyle D} = 1, 2, 3, 4$.
It is composed of two 
2-component fields that
transform the same way 
under rotations but
differently under Lorentz boosts.

\item[Momentum operator]
The hermitian 3-vector $\bos P$ is the
momentum operator.

\item[Spin operator]
The hermitian 3-vector $\bos S$ is the
spin operator.

\item[Orbital angular-momentum operator]
The hermitian 3-vector $\bos L$ is the
orbital angular-momentum operator.

\item[Angular-momentum operator]
The angular-momentum operator
is 
$\bos J ={} \bos L + \bos S$. 

\item[Ket $| \bos p, s \ra$]
The ket $|\bos p, s \ra$ is
an eigenvector of $\bos P$
with eigenvalue $\bos p$
and of $S_z$ with eigenvalue $s$,
so $\bos P |\bos p, s \ra 
= \bos p |\bos p, s \ra$ and
$S_z |\bos p, s \ra = s |\bos p, s \ra$.

\item[Rotations]
An active, right-handed rotation 
$R_{\bos\th}$ is a $3\by3$
orthogonal matrix that rotates 
3-vectors 
by $\th$ radians
about the
axis $\hat {\bos \th}$
in the right-handed way.
For instance, a right-handed
rotation by $\pi/2$ about
the $\hat {\bos y}$ axis
takes the 3-vector $\hat {\bos z}$ 
into ${} \hat {\bos x}$.

\item[Rotation operator]
The operator that represents
an active, right-handed 
rotation of $\th$ radians about
the axis $\hat {\bos \th}$ is the
unitary operator
$U(R_{\bos\th}) = 
\exp(-i \bos \th \cdot \bos J)$.
So the operator that represents
an active, right-handed
rotation of $\th$ radians about
the axis $\hat {\bos z}$ is the
unitary operator
$U(R_{\th \hat{\bos z}}) = 
\exp(-i \th J_z)$.

\item[Pauli matrices]
The Pauli matrices 
are the three $2\by 2$
hermitian matrices
\begin{equation}
\s_1 = \s_x ={} 
\Bigg(\begin{matrix}
0 & 1 \\
1 & 0
\end{matrix}\Bigg), \quad
\s_2 = \s_y ={}
\Bigg(\begin{matrix}
0 & - i \\
i & 0
\end{matrix}\Bigg),
\qand
\s_3 = \s_z ={}
\Bigg(\begin{matrix}
1 & 0 \\
0 & -1
\end{matrix}\Bigg).
\label{the Pauli matrices}
\end{equation}
They satisfy
$\s_i \s_j = \d_{ij} + i \ep_{ijk} \s_k$ .

\item[Commutators and anticommutators]
The commutator of two operators
$A$ and $B$ is $[A,B] \equiv AB - BA$;
their anticommutator  
is $\{A,B\} \equiv AB + BA$.

\item[Gamma matrices]
Dirac's gamma matrices
are any set of four $4\by4$ matrices
$\c^i$ that satisfy the anticommutation
relation
$ \{ \c^i, \c^k \}
\equiv \c^i \, \c^k + \c^k \, \c^i
={} 2 \eta^{ik}$.
The ones used in this paper are 
those of Weyl and 
Weinberg~\citep{WeinbergI216}
\begin{equation}
\c^0 ={} - i 
\Bigg(\begin{matrix}
0 & 1 \\
1 & 0 
\end{matrix}\Bigg)
\qand
\c^i ={} - i
\Bigg(\begin{matrix}
0 & \s^i \\
- \s^i & 0
\end{matrix}\Bigg).
\label{the gamma matrices}
\end{equation}
They have an extra factor of $i$
because I use the metric
$(-1, +1, +1, +1)$.
Every nonsingular $4\by4$ matrix
$S$ yields another set
of gamma matrices
$\c^{\prime i} = S \c^i S^{-1}$
that obey the condition
$ \{ \c'^i, \c'^k \} ={} 2 \eta^{ik}$.

\item[Equal-time anticommutation relations]
A Dirac field obeys the
equal-time anticommutation relations
$
\{ \psi_{\scriptscriptstyle D}(t, \bos x), 
\psi_{\scriptscriptstyle D'}(t, \bos y) \}
= 0$ and
$\{ \psi_{\scriptscriptstyle D}(t, \bos x), 
\psi^\dag_{\scriptscriptstyle D'}(t, \bos y) \}
={}
\d(\bos x - \bos y) \, \d_{\scriptscriptstyle D,
\scriptscriptstyle D'}$.
\end{description}

\section{Rotations of states of zero momentum and spin one-half 
\label{Rotations of states of zero momentum and spin one-half sec} 
}

The two states of a spin-one-half particle
of momentum zero transform 
under rotations like the 
spin-one-half systems of
nonrelativistic quantum 
mechanics as discussed
in, for example, chapter 3
of the book by 
Sakurai~\citep{Sakurai1994}\@.
This section reviews some details
that will be used in later sections.
\par
A spin-one-half particle
of momentum \( \bos p = \bos 0\) 
and spin \(s = \pm \thalf \)
in the \(z\) direction
is represented by a
state \( | \bos 0, s \ra \)
that is an eigenstate
of the momentum operator $\bos P$
and of the \(z\) component $S_z$
of the spin part $\bos S$ of the
angular-momentum operator
$\bos J = \bos L + \bos S$
with eigenvalues $\bos 0$ and $s$
\begin{equation}
\bos P | \bos 0, s \ra ={} \bos 0
\qand
S_z | \bos 0, s \ra ={} s | \bos 0, s \ra.
\end{equation}
The operator $\bos L$ is the
orbital part of the
angular-momentum operator
$\bos J = \bos L + \bos S$.
\par
The spin operator $S$
is represented 
in terms of the Pauli
matrices
\begin{equation}
\s_x ={} 
\Bigg(\begin{matrix}
0 & 1 \\
1 & 0
\end{matrix}\Bigg), \quad
\s_y ={}
\Bigg(\begin{matrix}
0 & - i \\
i & 0
\end{matrix}\Bigg),
\qand
\s_z ={}
\Bigg(\begin{matrix}
1 & 0 \\
0 & -1
\end{matrix}\Bigg).
\label{the Pauli matrices}
\end{equation}
as
\begin{equation}
\bos S = \frac{\hbar}{2} \bos \s
= \frac{\bos \s}{2} 
\quad \text{ or } \quad
\la s' | S_i | s \ra = \half  \lt( S_i \rt)_{s', s} .
\end{equation} 
\par
A right-handed, 
active rotation $R_{\bos\th}$
of $\th = |\bos\th|$ radians
about the axis \( \hat {\bos\th} \)
is represented
by the unitary operator
$e^{-i \bos \th \cdot \bos J} = {}
e^{-i \bos \th \cdot (\bos L + \bos S)}$.
When the 
momentum is zero,
this rotation 
leaves the momentum $\bos p = \bos 0$
unchanged.
So the total
angular-momentum operator
$\bos J ={} \bos L + \bos S$
and the spin angular-momentum operator
$\bos S$ have the same effect on
a state $ | \bos 0, s \ra$ of momentum zero.
They both rotate the state $ | \bos 0, s \ra$
to 
a linear combination of the
two spin states $ | \bos 0, \pm \thalf \ra$
\begin{equation}
\begin{split}
e^{-i \bos \th \cdot \bos J} |\bos 0, s\ra
={}&
e^{-i \bos \th \cdot \bos S} |\bos 0, s\ra
=
\sum_{s'=-\half}^\half
| \bos 0, s' \ra
\la s' | e^{-i \bos \th \cdot \bos S} | s\ra
\\
={}&
\sum_{s'=-\half}^\half
\lt[ 
e^{- i \bos \th \cdot \bos \s/2} 
\rt]_{s' s} \, | \bos 0, s' \ra 
= \sum_{s'=-\half}^\half
D_{s' s} (R_{\bos \th})
\, | \bos 0, s' \ra 
\label{rotation of states final}
\end{split}
\end{equation}
in which
\begin{equation}
I = {}
\sum_{s'=-\half}^\half
| s' \ra
\la s' |
\end{equation}
is the identity operator
and
\begin{equation}
D_{s' s}(R_{\bos \th}) ={} 
\la s' | e^{-i \bos \th \cdot \bos S} | s\ra
= \lt[ 
e^{- i \bos \th \cdot \bos \s/2} 
\rt]_{s' s}
\end{equation}
is the $2\by 2$ unitary matrix that represents
the rotation $R_{\bos\th}$.
\par
The identity
\begin{equation}
\s_i \s_j = \d_{ij} + i \sum_{k=1}^3\ep_{ijk} \s_k
\end{equation}
implies that
$ \lt( \bos \th \cdot \bos \s \rt)^2 
= \bos \th \cdot \bos \th $
which one may use to show that
the $SU(2)$ matrix 
$D_{s's}(R_{\bos \th)} = 
\lt[e^{-i \bos \th \cdot \bos \s/2}\rt]_{s' s}$
is a trigonometric combination 
of the Pauli matrices 
\begin{equation}    
D_{s' s}(R_{\bos \th}) ={} 
\d_{s' s} \, \cos (\th/2) 
- i (\hat{\bos \th} \cdot \bos \s)_{s' s} 
\, \sin (\th/2) .
\label {matrix D_z}
\end{equation} 

\par
For example, a rotation about 
the $\hat{\bos z}$ axis
by angle $\th$ is represented 
by the $2\by2$ matrix
\begin{equation}
D(R_{\th \hat{\bos z}}) ={}
e^{-i \th \s_z/2}
= I \, \cos(\th/2)  - i \s_z \, \sin(\th/2)
={} 
\begin{pmatrix}
e^{-i \th/2}  & 0 \\
0 &  e^{i \th/2}
\end{pmatrix} ,
\label{a rotation by theta about z axis}
\end{equation}
and so
\begin{equation}
e^{-i \th J_z} | \bos 0, s \ra
={}
e^{-i \th S_z} | \bos 0, s \ra
={} e^{-i \th s} | \bos 0, s \ra.
\label{z rotation of state 0,s}
\end{equation}
Similarly, a rotation by $\pi$
about the $y$ axis is represented
by the $2\by2$ matrix
\begin{equation}
D(R_{\pi \hat {\bos y}}) 
={} e^{-i \pi \s_y/2}
= I \, \cos(\pi/2)  - i \s_y \, \sin(\pi/2)
={} - i \s_y = 
\begin{pmatrix}
0 & -1 \\
1 & 0 
\end{pmatrix} ,
\label{a rotation by pi about y axis}
\end{equation}
and so
\begin{equation}
\begin{split}
e^{-i \pi J_y} | \bos 0, \thalf \ra
={}&
e^{-i \pi S_y} | \bos 0, \thalf \ra
={} | \bos 0, - \thalf \ra
\\
e^{-i \pi J_y} | \bos 0, \thalf \ra
={}&
e^{-i \pi S_y} | \bos 0, - \thalf \ra
={} - | \bos 0, \thalf \ra ,
\end{split}
\end{equation}
while 
a rotation by $\pi$
about the $x$ axis is represented
by the $2\by2$ matrix
\begin{equation}
D(R_{\pi \hat {\bos x}}) 
={} e^{-i \pi \s_x/2}
= I \, \cos(\pi/2)  - i \s_x \, \sin(\pi/2)
={} - i \s_x = 
\begin{pmatrix}
0 & -i \\
-i & 0 
\end{pmatrix} 
\label{a rotation by pi about x axis}
\end{equation}
which implies that
\begin{equation}
\begin{split}
e^{-i \pi J_x} | \bos 0, \thalf \ra
={}&
e^{-i \pi S_x} | \bos 0, \thalf \ra
={} - i| \bos 0, - \thalf \ra
\\
e^{-i \pi J_x} | \bos 0, \thalf \ra
={}&
e^{-i \pi S_x} | \bos 0, - \thalf \ra
={} -i | \bos 0, \thalf \ra .
\end{split}
\end{equation}

\section{Rotations of spin-one-half creation and annihilation operators of momentum zero
\label{Rotations of spin-one-half creation and annihilation operators of momentum zero sec}}

The creation and annihilation operators
of a spin-one-half particle of momentum zero
transform under rotations like the
two states of the spin-one-half
systems reviewed in 
Section~\ref{Rotations of states of zero momentum and spin one-half sec}.
\par
The state $ | \bos 0, s \ra $  
of a spin-one-half particle
of momentum zero and spin
$s = \pm \thalf$
is formed from 
the vacuum state \( | 0 \ra \)
by the creation operator \( a^\dag(\bos 0, s) \)
\begin{equation}
 a^\dag(\bos 0, s) \, | 0 \ra 
 ={} | \bos 0, s \ra .
 \label {adag 0 = 0s}
\end{equation}
Similarly, the antiparticle 
creation operator \( a^\dag_c(\bos 0, s) \)
adds an antiparticle of momentum
\( \bos 0 \) and spin $s$ to the vacuum state
\begin{equation}
 a^\dag_c(\bos 0, s) \, | 0 \ra 
 ={} | \bos 0, s \ra_c .
 \label {adag 0 = 0s}
\end{equation}

We've seen (\ref{z rotation of state 0,s}) that
a right-handed 
rotation about the $\hat{\bos z}$ axis
by angle $\th$ takes the state
$ | \bos 0, s \ra$ to 
$ e^{-i s \th} \, 
| \bos 0, s \ra $,
so we find
\begin{equation}
e^{-i \th J_z} \,  a^\dag(\bos 0, s) \, | 0 \ra 
 ={} e^{-i \th J_z} \,  | \bos 0, s \ra
 = e^{-i s \th} \,  | \bos 0, s \ra
 = e^{-i s \th} \, 
 a^\dag( \bos 0, s) \, | 0 \ra
\label {z rotation of state}
\end{equation}
and similarly
\begin{equation}
e^{-i \th J_z} \,  a^\dag_c(\bos 0, s) \, | 0 \ra 
 ={} e^{-i \th J_z} \,  | \bos 0, s \ra_c
 = e^{-i s \th} \,  | \bos 0, s \ra_c
 = e^{-i s \th} \, 
 a^\dag_c( \bos 0, s) \, | 0 \ra_c .
\label {z rotation of state c}
\end{equation}
In the flat space-time
of quantum field theory, 
the vacuum is invariant under rotations,
so
\begin{equation}
e^{i \th J_z} \, | 0 \ra = | 0 \ra .
\end{equation}
Thus we can rewrite equations 
(\ref{z rotation of state} and 
\ref{z rotation of state c}) as
\begin{equation}
\begin{split}
e^{-i \th J_z} \,  a^\dag(\bos 0, s) \, 
e^{i \th J_z} \, | 0 \ra 
 ={}& e^{-i s \th} \, 
 a^\dag(  \bos 0, s) \, | 0 \ra
 \\
e^{-i \th J_z} \,  a^\dag_c(\bos 0, s) \, 
e^{i \th J_z} \, | 0 \ra 
 ={}& e^{-i s \th} \, 
 a^\dag_c(  \bos 0, s) \, | 0 \ra .
 \label {a dag vacuum case}
 \end{split}
\end{equation}
Replacing 
the vacuum state $|0\ra$
by an arbitrary state $|\psi\ra$
and repeating the last few steps,
we see that
a $z$ rotation changes
the creation operators $a^\dag(\bos 0,s)$
and $a^\dag_c(\bos 0,s)$ to
\begin{equation}
\begin{split}
e^{-i \th J_z} \,  a^\dag(\bos 0, s) \, 
e^{i \th J_z}
 ={}& e^{-i s \th} \, 
 a^\dag(  \bos 0, s)
 \\
e^{-i \th J_z} \,  a^\dag_c(\bos 0, s) \, 
e^{i \th J_z}
 ={}& e^{-i s \th} \, 
 a^\dag_c(  \bos 0, s)  .
\label {creation operator goes as}
\end{split}
\end{equation}
The adjoints of these equations are
\begin{equation}
   \begin{split}
e^{-i \th J_z} a( \bos 0, s)
e^{i \th J_z} 
={}& 
e^{i s \th } a( \bos 0, s) 
\\
e^{-i \th J_z} a_c( \bos 0, s)
e^{i \th J_z} 
={}& 
e^{i s \th } a_c( \bos 0, s) 
\label {annihilation operator goes as}
\end{split}
\end{equation}
Under rotations,
creation and annihilation operators
transform with opposite phases.
That's why the \(u(\bos 0, s)\) spinors
that multiply annihilation operators
are and different from 
the \(v(\bos 0, s)\) spinors 
that multiply creation operators.
\par
If instead of the $z$ axis,
the rotation is 
about an arbitrary axis $\hat {\bos \th}$,
then the equations that replace 
(\ref{creation operator goes as}
and
\ref{annihilation operator goes as})
are
\begin{equation}
\begin{split}
e^{-i \bos \th \cdot \bos J} a(\bos 0, s)
e^{i \bos\th \cdot \bos J} 
={}& 
\sum_{s'}
D^*_{s' s}(R_{\bos \th}) \,\,
a(\bos 0, s')
=
\sum_{s'}
D^{-1}_{s s'}(R_{\bos \th}) \,\,
a(\bos 0, s')
\\
e^{-i \th \cdot \bos J} a^\dag_c(\bos 0, s)
\, e^{i \bos\th \cdot \bos J} 
={}& 
\sum_{s'}
D_{s' s}(R_{\bos \th}) \,
a^\dag_c(\bos 0, s') 
=
\sum_{s'}
D^{* -1}_{s s'}(R_{\bos \th}) \,
a^\dag_c(\bos 0, s') .
\label{how a and adag go under rotations}
\end{split}
\end{equation}
In particular, 
a rotation by $\pi$ about the $y$ axis
changes $a(\bos 0, s)$ and $a^\dag_c(\bos 0, s)$
to
\begin{equation}
\begin{split}
e^{-i \pi J_y} a(\bos 0, s)
e^{i \pi J_y} 
={}& 
\sum_{s'}
D^*_{s' s}(R_{\pi \hat{\bos y}}) \,\,
a(\bos 0, s')
=
\sum_{s'}
D^{-1}_{s s'}(R_{\pi \hat{\bos y}}) \,\,
a(\bos 0, s')
\\
e^{-i \pi J_y} a^\dag_c(\bos 0, s)
\, e^{i \pi J_y} 
={}& 
\sum_{s'}
D_{s' s}(R_{\pi \hat{\bos y}}) \,
a^\dag_c(\bos 0, s') 
=
\sum_{s'}
D^{* -1}_{s s'}(R_{\pi \hat{\bos y}}) \,
a^\dag_c(\bos 0, s') .
\label{rotation about y axis a adag}
\end{split}
\end{equation}
That is,
\begin{equation}
\begin{split}
e^{-i \pi J_y} 
\begin{pmatrix}
a(\bos 0, \thalf) \\
a(\bos 0, -\thalf)
\end{pmatrix}
e^{i \pi J_y} 
={}&
\begin{pmatrix}
0 & 1\\
-1 & 0
\end{pmatrix}
\begin{pmatrix}
a(\bos 0, \thalf) \\
a(\bos 0, -\thalf)
\end{pmatrix}
=
\begin{pmatrix}
a(\bos 0, -\thalf) \\
- a(\bos 0, \thalf)
\end{pmatrix}
\label{y rotation of a's}
\end{split}
\end{equation}
and
\begin{equation}
\begin{split}
e^{-i \pi J_y} 
\begin{pmatrix}
a^\dag_c(\bos 0, \thalf) \\
a^\dag_c(\bos 0, -\thalf)
\end{pmatrix}
e^{i \pi J_y} 
={}&
\begin{pmatrix}
0 & 1\\
-1 & 0
\end{pmatrix}
\begin{pmatrix}
a^\dag_c(\bos 0, \thalf) \\
a^\dag_c(\bos 0, -\thalf)
\end{pmatrix}
=
\begin{pmatrix}
a^\dag_c(\bos 0, -\thalf) \\
- a^\dag_c(\bos 0, \thalf)
\end{pmatrix} .
\label{rotation about y axis adag_c}
\end{split}
\end{equation}
And 
a rotation by $\pi$ about the $x$ axis
changes $a(\bos 0, s)$ and $a^\dag_c(\bos 0, s)$
to
\begin{equation}
\begin{split}
e^{-i \pi J_x} a(\bos 0, s)
e^{i \pi J_x} 
={}& 
\sum_{s'}
D^*_{s' s}(R_{\pi \hat{\bos x}}) \,\,
a(\bos 0, s')
=
\sum_{s'}
D^{-1}_{s s'}(R_{\pi \hat{\bos x}}) \,\,
a(\bos 0, s')
\\
e^{-i \pi J_x} a^\dag_c(\bos 0, s)
\, e^{i \pi J_x} 
={}& 
\sum_{s'}
D_{s' s}(R_{\pi \hat{\bos x}}) \,
a^\dag_c(\bos 0, s') 
=
\sum_{s'}
D^{* -1}_{s s'}(R_{\pi \hat{\bos x}}) \,
a^\dag_c(\bos 0, s') .
\end{split}
\end{equation}
That is,
\begin{equation}
\begin{split}
e^{-i \pi J_x} 
\begin{pmatrix}
a(\bos 0, \thalf) \\
a(\bos 0, -\thalf)
\end{pmatrix}
e^{i \pi J_x} 
={}&
\begin{pmatrix}
0 & i\\
i & 0
\end{pmatrix}
\begin{pmatrix}
a(\bos 0, \thalf) \\
a(\bos 0, -\thalf)
\end{pmatrix}
=
\begin{pmatrix}
i a(\bos 0, -\thalf) \\
i a(\bos 0, \thalf)
\end{pmatrix}
\end{split}
\end{equation}
and
\begin{equation}
\begin{split}
e^{-i \pi J_x} 
\begin{pmatrix}
a^\dag(\bos 0, \thalf) \\
a^\dag(\bos 0, -\thalf)
\end{pmatrix}
e^{i \pi J_x} 
={}&
\begin{pmatrix}
0 & -i\\
-i & 0
\end{pmatrix}
\begin{pmatrix}
a^\dag(\bos 0, \thalf) \\
a^\dag(\bos 0, -\thalf)
\end{pmatrix}
=
\begin{pmatrix}
- i a^\dag(\bos 0, -\thalf) \\
- i a^\dag(\bos 0, \thalf).
\end{pmatrix}
\end{split}
\end{equation}

\section{Rotations of two-component spin-one-half fields
\label{Rotations of two-component spin-one-half fields sec}}

A Dirac field is made of two
2-component spin-one-half fields
that transform the same way 
under rotations but differently
under Lorentz boosts.
This Section explains how these 
2-component spin-one-half fields
transform under rotations
and derives formulas
for their 2-component spinors
at momentum zero.
\par
Like the upper or lower 
two components of
the Fourier expansion 
(\ref{4 component Dirac field I})
of a four-component
spin-one-half field,
the Fourier expansion 
of a two-component
spin-one-half field is 
\begin{equation}
   \begin{split}
\psi_\a(x) = {}&
\sum_{s=\pm 1/2}
\int \frac{d^3p}{(2 \pi)^{3/2}} 
\lt[
u_\a(\bos p, s) \, e^{ip\cdot x} \, a(\bos p,s)
+
v_\a(\bos p, s) \, e^{-ip\cdot x} 
\, a_c^\dag(\bos p, s) \rt]
\label {massive Dirac field 2}
   \end{split}
\end{equation}
in which $\a = 1, 2$ and  
$p \cdot x ={} \bos p \cdot \bos x
- p^0 t$ with $p^0 = {} 
\sqrt{\bos p^2 + m^2}$.
\par
Under a rotation $R_{\bos\th}$
represented by the matrix
$D(R_{\bos \th}) = {}
e^{-i \bos \th \cdot \bos \s/2}$
of equation~\ref{matrix D_z}, 
a 2-component 
spin-one-half field $\psi_\a(x)$
transforms as
\begin{equation}
\begin{split}
U(R_{\bos\th}) \psi_\a(x) U^{-1}(R_{\bos\th}) 
={}&
e^{-i \pi J_x} 
\psi_\a(x)
e^{i \pi J_x} 
=
\sum_{\b = 1}^2
D_{\a\b}(R^{-1}_{\bos\th})
\,  \psi_\b(R_{\bos\th}x)
\\
={}&
\sum_{\b = 1}^2
\Big(e^{i \thalf \bos \th \cdot \bos \s }\Big)_{\a\b} 
\psi_\b(R_{\bos\th}x).
\label{rotation of spin 1/2 field}
\end{split}
\end{equation}
This equation may be taken as part of the
definition of a spin-one-half field,
another part being
how the field transforms 
(\ref{Lorentz transformation}) under
Lorentz transformations that are not
simple rotations.
But because the effects of rotations 
can be confusing,
I will sketch one aspect of
this equation 
(\ref{rotation of spin 1/2 field})
by writing its mean value
in some state $| W \ra$ of the world
as
\begin{equation}
\la W | U(R_{\bos\th}) \psi(x) 
U^{-1}(R_{\bos\th}) | W \ra
={}
\la R^{-1} W |  \psi(x) 
 | R^{-1} W \ra
={}
D(R^{-1}) \la W | \psi(Rx) | W \ra.
\end{equation}
Apart from the matrix
$D(R^{-1})$, this equation
says that the mean value
of the field at $x$ in a world
rotated by $R^{-1}$ is the same
as its mean value at $Rx$ in
an unrotated world.
But the field $\psi$ is a vector
whose mean value depends upon
the world.  So in a world
rotated by $R^{-1}$, its mean value
is rotated by $R^{-1}$.
\par
The matrix $D(R_{\th \hat {\bos z}})$ 
that represents a rotation
about the $z$ axis
(\ref{a rotation by theta about z axis})
is diagonal
because $\s_z$ is diagonal
\begin{equation}
D(R_{\th \hat {\bos z}}) = {} 
e^{-i \th \s_z} =
\begin{pmatrix}
e^{-i \th} & 0 \\
0 & e^{i \th} 
\end{pmatrix}.
\end{equation}
So under a rotation about the 
$\bos \th = \th \hat {\bos z}$ axis,
the field $\psi_\a(x)$ transforms as 
\begin{equation}
\begin{split}
U(R_{\th \hat {\bos z}}) \psi_\a(x) U^{-1}(R_{\th \hat {\bos z}}) 
={}&
e^{-i  \th J_z} 
\psi_\a(x)
e^{i \th J_z} 
=
\sum_{\b = 1}^2
\exp(i \thalf \th \s_z )_{\a\b} 
\psi_\b(R_{\th \hat {\bos z}}x) .
\label {z rotation of spin 1/2 field}
\end{split}
\end{equation}
\par
We are after the spinors at momentum zero,
so we need pay attention only to the
momentum-zero part $\psi^0_\a(x)(x)$ 
of the field.
We also can distinguish between the
annihilator part $\psi^{(0,+)}_\a(x)$
and the creator part
$\psi^{(0,-)}_\a(x)$.
Apart from the factors of $2\pi$,
the momentum-zero, 
annihilator part $\psi^{(0,+)}_\a(x)$ 
at $x=0$ is
\begin{equation}
\begin{pmatrix}
\psi^{(0,+)}_1(0) \\
\psi^{(0,+)}_2(0)
\end{pmatrix} ={}
\begin{pmatrix}
u_1(0,\thalf) \\
u_2(0,\thalf)
\end{pmatrix} a(0, \thalf)
\, + \,
\begin{pmatrix}
u_1(0,-\thalf) \\
u_2(0,-\thalf)
\end{pmatrix} a(0, -\thalf).
\label{momentum-zero, annihilator part}
\end{equation}
The formula 
(\ref{rotation of spin 1/2 field})
for how the field $\psi_\a(x)$
transforms under the rotation
$R_{\bos\th}$ simplifies
for the momentum-zero, 
annihilator part $\psi^{(0,+)}_\a(x)$ 
at $x=0$ to
\begin{equation}
e^{i \thalf \bos \th \cdot \bos \s }
\, \psi^{(0,+)}(0)
={}
e^{-i \thalf \bos \th \cdot \bos \s } 
\psi^{(0,+)}(0)
e^{i \thalf \bos \th \cdot \bos \s } .
\label{rotation of annihilators}
\end{equation}
For a rotation
about the $z$ axis,
this is more explicitly
\begin{equation}
\begin{split}
\begin{pmatrix}
e^{i\th/2} & 0 \\
0 & e^{-i\th/2}
\end{pmatrix}
&
\lt[
\begin{pmatrix}
u_1(0,\thalf) \\
u_2(0,\thalf)
\end{pmatrix} 
a(0, \thalf)
\, + \,
\begin{pmatrix}
u_1(0,-\thalf) \\
u_2(0,-\thalf)
\end{pmatrix} 
a(0, -\thalf)
\rt]
\\
={}&
\begin{pmatrix}
e^{i\th/2} \, u_1(0,\thalf) \\
e^{-i\th/2} \, u_2(0,\thalf)
\end{pmatrix} 
a(0, \thalf)
\, + \,
\begin{pmatrix}
e^{i\th/2} \, u_1(0,-\thalf) \\
e^{-i\th/2} \, u_2(0,-\thalf)
\end{pmatrix} 
a(0, -\thalf)
\\
={}&
\begin{pmatrix}
u_1(0,\thalf) \\
u_2(0,\thalf)
\end{pmatrix} 
e^{i\th/2} \, a(0, \thalf)
\, + \,
\begin{pmatrix}
u_1(0,-\thalf) \\
u_2(0,-\thalf)
\end{pmatrix} 
e^{-i\th/2} \, a(0, -\thalf).
\label{z rotation of u}
\end{split}
\end{equation}
We see that the 
momentum-zero 2-spinors are
\begin{equation}
u(0,\thalf) ={} 
c_+ \begin{pmatrix}
1 \\ 0 
\end{pmatrix}
\qand \, 
u(0,-\thalf) ={} 
c_-  \begin{pmatrix}
0 \\ 1 
\end{pmatrix}
\end{equation}
as one would have expected
without the present derivation.
\par
Apart from the factors of $2\pi$,
the momentum-zero, 
creator part $\psi^{(0,-)}_\a(x)$ 
at $x=0$ is
\begin{equation}
\begin{pmatrix}
\psi^{(0,-)}_1(0) \\
\psi^{(0,-)}_2(0)
\end{pmatrix} ={}
\begin{pmatrix}
v_1(0,\thalf) \\
v_2(0,\thalf)
\end{pmatrix} a^\dag_c(0, \thalf)
\, + \,
\begin{pmatrix}
v_1(0,-\thalf) \\
v_2(0,-\thalf)
\end{pmatrix} a^\dag_c(0, -\thalf).
\label{momentum-zero, annihilator part}
\end{equation}
\par
The formula 
(\ref{rotation of spin 1/2 field})
for how the field $\psi_\a(x)$
transforms under the rotation
$R_{\bos\th}$ simplifies
for the momentum-zero, 
creator part $\psi^{(0,-)}_\a(x)$ 
at $x=0$ to
\begin{equation}
e^{i \thalf \bos \th \cdot \bos \s }
\, \psi^{(0,-)}(0)
={}
e^{-i \bos \th \cdot \bos J} 
\psi^{(0,-)}(0)
e^{i \bos \th \cdot \bos J}.
\label{rotation of creators}
\end{equation}
For a rotation
about the $z$ axis,
this is more explicitly
\begin{equation}
\begin{split}
\begin{pmatrix}
e^{i\th/2} & 0 \\
0 & e^{-i\th/2}
\end{pmatrix}
&
\lt[
\begin{pmatrix}
v_1(0,\thalf) \\
v_2(0,\thalf)
\end{pmatrix} 
a^\dag_c(0, \thalf)
\, + \,
\begin{pmatrix}
v_1(0,-\thalf) \\
v_2(0,-\thalf)
\end{pmatrix} 
a^\dag_c(0, -\thalf)
\rt]
\\
={}&
\begin{pmatrix}
e^{i\th/2} \, v_1(0,\thalf) \\
e^{-i\th/2} \, v_2(0,\thalf)
\end{pmatrix} 
a^\dag_c(0, \thalf)
\, + \,
\begin{pmatrix}
e^{i\th/2} \, v_1(0,-\thalf) \\
e^{-i\th/2} \, v_2(0,-\thalf)
\end{pmatrix} 
a^\dag_c(0, -\thalf)
\\
={}&
\begin{pmatrix}
v_1(0,\thalf) \\
v_2(0,\thalf)
\end{pmatrix} 
e^{-i\th/2} \, a^\dag_c(0, \thalf)
\,\, + \,\,
\begin{pmatrix}
v_1(0,-\thalf) \\
v_2(0,-\thalf)
\end{pmatrix} 
e^{i\th/2} \, a^\dag_c(0, -\thalf).
\end{split}
\end{equation}
We see that the 
momentum-zero 2-spinors are
\begin{equation}
v(0,\thalf) ={} 
d_+ \begin{pmatrix}
0 \\ 1 
\end{pmatrix}
\qand \,
v(0,-\thalf) ={} 
d_-  \begin{pmatrix}
1 \\ 0 
\end{pmatrix}
\end{equation}
which may surprise us.
These formulas
are missed in three
leading textbooks~\citep{Ryder:1985wq,Kaku:1993ym,Schwartz:p189}.

\par
We use our knowledge 
(\ref{y rotation of a's}) of
how annihilation operators
respond to a rotation 
(\ref{a rotation by pi about y axis})
by angle $\pi$
about the $y$ axis.
We use the matrix
(\ref{a rotation by pi about y axis})
that represents 
a rotation by $\pi$ about $y$ axis
\begin{equation}
e^{i\pi \s_y/2}
={}
\begin{pmatrix}
0 & 1 \\
-1 & 0
\end{pmatrix},
\end{equation}
and our rule (\ref{rotation of annihilators})
for how the annihilator part
of the momentum-zero field
transforms 
\begin{equation}
e^{-i \pi J_y/2} 
\psi^{(0,+)}(0)
e^{i \pi J_y/2}
={}
e^{i \pi \s_y/2 }
\, \psi^{(0,+)}(0),
\label{rotation of annihilators}
\end{equation}
which in more detail 
(\ref{y rotation of a's}) is
\begin{equation}
\begin{split}
e^{-i \pi J_y} 
\begin{pmatrix}
a(\bos 0, \thalf) \\
a(\bos 0, -\thalf)
\end{pmatrix}
e^{i \pi J_y} 
={}&
\begin{pmatrix}
0 & 1\\
-1 & 0
\end{pmatrix}
\begin{pmatrix}
a(\bos 0, \thalf) \\
a(\bos 0, -\thalf)
\end{pmatrix}
=
\begin{pmatrix}
a(\bos 0, -\thalf) \\
- a(\bos 0, \thalf)
\end{pmatrix}
\end{split}
\end{equation}
to infer that
\begin{equation}
\begin{split}
\begin{pmatrix}
0 & 1 \\
-1 & 0
\end{pmatrix}
&
\lt[
c_+ \begin{pmatrix}
1 \\
0
\end{pmatrix} 
a(0, \thalf)
+
c_- \begin{pmatrix}
0 \\
1
\end{pmatrix} 
a(0, -\thalf)
\rt]
\\
={}&
c_+ \begin{pmatrix}
0 \\
- 1
\end{pmatrix} 
a(0, \thalf)
+
c_- \begin{pmatrix}
1 \\
0
\end{pmatrix} 
a(0, -\thalf)
\\
={}&
e^{-i\pi J_y}
\lt[
c_+ \begin{pmatrix}
1 \\
0
\end{pmatrix} 
a(0, \thalf)
+
c_- \begin{pmatrix}
0 \\
1
\end{pmatrix} 
a(0, -\thalf)
\rt] e^{i\pi J_y}
\\
={}&
c_+ \begin{pmatrix}
1 \\
0
\end{pmatrix} 
a(0, -\thalf)
-
c_- \begin{pmatrix}
0 \\
1
\end{pmatrix} 
a(0, \thalf) .
\label{z rotation of u}
\end{split}
\end{equation}
So now we know that
$c_- ={} c_+$,
as may have been anticipated.
\par
Under the same $y$ rotation by $\pi$,
the creation operators transform as
(\ref{rotation about y axis adag_c})
\begin{equation}
\begin{split}
e^{-i \pi J_y} 
\begin{pmatrix}
a^\dag_c(\bos 0, \thalf) \\
a^\dag_c(\bos 0, -\thalf)
\end{pmatrix}
e^{i \pi J_y} 
={}&
\begin{pmatrix}
0 & 1\\
-1 & 0
\end{pmatrix}
\begin{pmatrix}
a^\dag_c(\bos 0, \thalf) \\
a^\dag_c(\bos 0, -\thalf)
\end{pmatrix}
=
\begin{pmatrix}
a^\dag_c(\bos 0, -\thalf) \\
- a^\dag_c(\bos 0, \thalf)
\end{pmatrix} ,
\end{split}
\end{equation}
and so the formula
(\ref{rotation of creators})
for how the zero-momentum part
of the creator field transforms
under rotations gives us
\begin{equation}
\begin{split}
\begin{pmatrix}
0 & 1 \\
-1 & 0
\end{pmatrix}
&
\lt[
d_+ \begin{pmatrix}
0 \\ 1 
\end{pmatrix}
a^\dag_c(0, \thalf)
+
d_-  \begin{pmatrix}
1 \\ 0 
\end{pmatrix}
a^\dag_c(0, -\thalf)
\rt]
\\
={}&
d_+ \begin{pmatrix}
1 \\ 0 
\end{pmatrix}
a^\dag_c(0, \thalf)
+
d_-  \begin{pmatrix}
0 \\ -1 
\end{pmatrix}
a^\dag_c(0, -\thalf)
\\
={}&
d_+ \begin{pmatrix}
0 \\ 1 
\end{pmatrix}
a^\dag_c(0, -\thalf)
-
d_-  \begin{pmatrix}
1 \\ 0 
\end{pmatrix}
a^\dag_c(0, \thalf) .
\end{split}
\end{equation}
So now we know that
$d_+ = - d_-$.
\par
Does the $x$ rotation give the 
same answers?
We use the equations
\begin{equation}
e^{i\pi \s_x/2}
={}
\begin{pmatrix}
0 & i \\
i & 0
\end{pmatrix},
\end{equation}
and
\begin{equation}
e^{i \pi \s_x/2 }
\, \psi^{(0,+)}(0)
={}
e^{-i \pi J_x/2} 
\psi^{(0,+)}(0)
e^{i \pi J_x/2},
\label{rotation of annihilators}
\end{equation}
and 
\begin{equation}
\begin{split}
e^{-i \pi J_x} 
\begin{pmatrix}
a(\bos 0, \thalf) \\
a(\bos 0, -\thalf)
\end{pmatrix}
e^{i \pi J_x} 
={}&
\begin{pmatrix}
0 & i\\
i & 0
\end{pmatrix}
\begin{pmatrix}
a(\bos 0, \thalf) \\
a(\bos 0, -\thalf)
\end{pmatrix}
=
\begin{pmatrix}
i a(\bos 0, -\thalf) \\
i a(\bos 0, \thalf)
\end{pmatrix}
\end{split}
\end{equation}
to infer that
\begin{equation}
\begin{split}
\begin{pmatrix}
0 & i \\
i & 0
\end{pmatrix}
&
\lt[
c_+ \begin{pmatrix}
0 \\
1
\end{pmatrix} 
a(0, \thalf)
+
c_- \begin{pmatrix}
1 \\
0
\end{pmatrix} 
a(0, -\thalf)
\rt]
\\
={}&
c_+ \begin{pmatrix}
i \\
0
\end{pmatrix} 
a(0, \thalf)
+
c_- \begin{pmatrix}
0 \\
i
\end{pmatrix} 
a(0, -\thalf)
\\
={}&
e^{-i\pi J_x}
\lt[
c_+ \begin{pmatrix}
0 \\
1
\end{pmatrix} 
a(0, \thalf)
+
c_- \begin{pmatrix}
1 \\
0
\end{pmatrix} 
a(0, -\thalf)
\rt] e^{i\pi J_x}
\\
={}&
c_+ \begin{pmatrix}
0 \\
i
\end{pmatrix} 
a(0, -\thalf)
+
c_- \begin{pmatrix}
i \\
0
\end{pmatrix} 
a(0, \thalf)
\label{z rotation of u}
\end{split}
\end{equation}
which again tells us that
$c_- ={} c_+$.
\par
The creation versions
of these equations are
\begin{equation}
\begin{split}
e^{-i \pi J_x} 
\begin{pmatrix}
a^\dag_c(\bos 0, \thalf) \\
a^\dag_c(\bos 0, -\thalf)
\end{pmatrix}
e^{i \pi J_x} 
={}&
\begin{pmatrix}
0 & -i\\
-i & 0
\end{pmatrix}
\begin{pmatrix}
a^\dag_c(\bos 0, \thalf) \\
a^\dag_c(\bos 0, -\thalf)
\end{pmatrix}
=
\begin{pmatrix}
- ia^\dag_c(\bos 0, -\thalf) \\
- ia^\dag_c(\bos 0, \thalf)
\end{pmatrix} .
\end{split}
\end{equation}
and
\begin{equation}
\begin{split}
\begin{pmatrix}
0 & i \\
i & 0
\end{pmatrix}
&
\lt[
d_+ \begin{pmatrix}
0 \\ 1 
\end{pmatrix}
a^\dag_c(0, \thalf)
+
d_-  \begin{pmatrix}
1 \\ 0 
\end{pmatrix}
a^\dag_c(0, -\thalf)
\rt]
\\
={}&
d_+ \begin{pmatrix}
i \\ 
0 
\end{pmatrix}
a^\dag_c(0, \thalf)
+
d_-  \begin{pmatrix}
0 \\ 
i 
\end{pmatrix}
a^\dag_c(0, -\thalf)
\\
={}&
d_+ \begin{pmatrix}
0 \\ 
-i 
\end{pmatrix}
a^\dag_c(0, -\thalf)
+
d_-  \begin{pmatrix}
-i \\ 
0 
\end{pmatrix}
a^\dag_c(0, \thalf)
\end{split}
\end{equation}
which again tell us that
$d_+ = - d_-$.
\par
So with $c = c_+ = c_-$,
and $d = d_+ = -d_-$,
our momentum-zero
spinors 
are
\begin{equation}
\begin{split}
u(0, \thalf) ={}& c \bpm 1 \\ 0 \epm
\qand
u(0, -\thalf) ={} c \bpm 0 \\ 1 \epm
\\
v(0, \thalf) ={}& d \bpm 0 \\ 1 \epm
\qand
v(0, -\thalf) ={} d \bpm -1 \\ 0 \epm.
\label{our spinors now are}
\end{split}
\end{equation}
We have derived these formulas 
for 2-component spinors from
the Pauli matrices and the $SU(2)$
representation of rotations that they
provide.  These formulas therefore 
are independent of the choice
(\ref{the gamma matrices}, 
\ref{the Weyl gamma matrices})
of gamma matrices\@.

\section{Four-component spinors at zero momentum}
\label{Four-component spinors at zero momentum sec}

In this Section, 
we will use the 2-component spinors
(\ref{our spinors now are})
and the
Dirac equation
in momentum space
at momentum zero
to derive formulas for
the 4-component spinors
$u(\bos 0,s)$ and $v(\bos 0,s)$
at momentum zero.
\par
We make 4-component
$u$ spinors by
putting two pairs 
of 2-component  
$u$ spinors (\ref{our spinors now are}) 
together
\begin{equation}
u(0, \thalf) ={} 
\bpm u_\ell(0, \thalf) \\ u_r(0, \thalf) \epm
= \bpm c_\ell \\ 0 \\ c_r \\ 0 \epm
\qand
u(0, -\thalf) ={} 
\bpm u_\ell(0, -\thalf) \\ u_r(0, -\thalf) \epm
= \bpm 0 \\ c_\ell\\ 0 \\ c_r \epm
\label{four-component Dirac u spinors}
\end{equation}
and by putting two pairs 
of 2-component  
$v$ spinors (\ref{our spinors now are}) 
together
\begin{equation}
v(0, \thalf) ={} 
\bpm v_\ell(0, \thalf) \\ v_r(0, \thalf) \epm
= \bpm 0 \\ d_\ell \\ 0 \\ d_r \epm
\qand
v(0, -\thalf) ={} 
\bpm v_\ell(0, -\thalf) \\ v_r(0, -\thalf) \epm
= \bpm - d_\ell \\ 0 \\ - d_r\\ 0 \epm.
\label{four-component Dirac v spinors}
\end{equation} 
in which  
$c_\ell$, $c_r$,
$d_\ell$, and $d_r$
are arbitrary constants.
Dirac's equation at momentum zero
will tell us that $c_\ell = c_r$
and that $v_\ell = {} - v_r$\@.
\par
The labels $\ell$ and $r$ 
refer to whether the spinors behave
as left-handed or right-handed 
2-component spinors 
under Lorentz 
boosts. 
How a Dirac field is made out of 
a 2-component left-handed field
and a 2-component right-handed field
is outlined in Section~\ref{Left-handed and right-handed spin-one-half fields sec}
and described more fully in
references~\citep{WeinbergIspinors}
and \citep{CahillCUP2DiracField}.
\par
To narrow down the range
of the constants $c_\ell$, $c_r$,
$d_\ell$, and $d_r$
in the 4-component spinors
(\ref{four-component Dirac u spinors}
and \ref{four-component Dirac v spinors}), 
we use Dirac's equation 
\begin{equation}
(\bos \c \cdot \grad - \c^0 \p_0 + m) 
\psi(x)
=
(\c^a \p_a + m) 
\psi(x) = 0
\label {the Dirac equation}
\end{equation}
and his 
gamma matrices (\ref{the gamma matrices})
\begin{equation}
\c^0 ={} - i 
\Bigg(\begin{matrix}
0 & 1 \\
1 & 0 
\end{matrix}\Bigg)
\qand
\c^i ={} - i
\Bigg(\begin{matrix}
0 & \s^i \\
- \s^i & 0
\end{matrix}\Bigg)
\label{the Weyl gamma matrices}
\end{equation}
in which the $\s_i$'s are the Pauli matrices 
(\ref{the Pauli matrices})
and the extra factors of $i$ ensure
that the anticommutator of 
two gamma matrices is
$\{ \c^a, \c^b \} = {} 2 \eta^{ab}$
in which the diagonal of $\eta$ is
$(-1, 1, 1, 1)$\@. 
\par
When the derivative $\p_a$
in the Dirac equation
acts on 
$u(\bos p,s) e^{ip\cdot x}$,
it becomes $i p_a$;
when it acts on 
$v(\bos p,s) e^{-ip\cdot x}$,
becomes $-i p_a$.
This is the second reason
why the spinors $u$
that multiply annihilation
operators are different from
those $v$ that multiply creation
operators.
\par
The Dirac equation 
in momentum space 
therefore
splits into one equation
for $u$ spinors and 
another for $v$ spinors
\begin{equation}
(i p_a \c^a + m) u(\bos p,s) = 0
\qand
(- i p_a \c^a + m) v(\bos p,s) = 0.
\label {equations for u, v spinors}
\end{equation}
At $p_a = (-m,0,0,0)$, 
these equations are
$(-i \c^0 + 1) u(\bos 0, s) = 0$
for $s = \pm \thalf$
\begin{equation}
\begin{pmatrix}
1 & 0 & -1 & 0 \\ 
0 & 1 & 0 & -1 \\
-1 & 0 & 1 & 0 \\
0 & -1 & 0 & 1 
\end{pmatrix}
\bpm c_\ell \\ 0 \\ c_r \\ 0 \epm
= 0
\qand
\begin{pmatrix}
1 & 0 & -1 & 0 \\ 
0 & 1 & 0 & -1 \\
-1 & 0 & 1 & 0 \\
0 & -1 & 0 & 1 
\end{pmatrix}
\bpm 0 \\ c_\ell\\ 0 \\ c_r \epm
= 0
\end{equation}
and 
$(i \c^0 + 1) v(\bos 0, s) = 0$
for $s = \pm \thalf$
\begin{equation}
\begin{pmatrix}
1 & 0 & 1 & 0 \\ 
0 & 1 & 0 & 1 \\
1 & 0 & 1 & 0 \\
0 & 1 & 0 & 1 
\end{pmatrix}
\bpm 0 \\ d_\ell\\ 0 \\ d_r \epm
= 0
\qand
\begin{pmatrix}
1 & 0 & 1 & 0 \\ 
0 & 1 & 0 & 1 \\
1 & 0 & 1 & 0 \\
0 & 1 & 0 & 1 
\end{pmatrix}
\bpm -d_\ell \\ 0 \\ -d_r \\ 0 \epm
= 0.
\label{gamma0 and v spinors}
\end{equation}
They require that 
$c_\ell = c_r$ and that
$d_\ell = {} - d_r$.
So with 
$c_\ell = c_r = c$
and $d_\ell = - d_r = d$,
our spinors are 
\begin{equation}
u(0, \thalf) ={} c 
\bpm 1 \\ 0 \\ 1 \\ 0 \epm,
\quad
u(0, -\thalf) ={} c 
\bpm 0 \\ 1 \\ 0 \\ 1 \epm,
\quad
v(0, \thalf) ={} d
\begin{pmatrix}
0 \\ 1 \\ 0 \\ -1
\end{pmatrix},
\quad
v(0, {}-\thalf) ={} d
\begin{pmatrix}
-1 \\ 0 \\ 1 \\ 0
\end{pmatrix}.
\end{equation}
Choosing the arbitrary relative phases
of creation and annihilation
operators and the arbitrary phase
of the field and normalizing the
spinors, we get 
the zero-momentum 
spinors derived by
Weinberg~\citep{WeinbergIp=0spinors}
\begin{equation}
\begin{split}
u(\bos0,  \thalf) ={} &
\frac{1}{\sqrt{2}}
 \begin{pmatrix}
1 \\
0 \\
1 \\
0 \\
\end{pmatrix} , \qquad
u(\bos0, -\thalf) = {} 
\frac{1}{\sqrt{2} }
\begin{pmatrix}
0 \\ 1 \\ 0 \\ 1
\end{pmatrix} 
\\
v(\bos0, \thalf) ={}&
\frac{1}{\sqrt{2}}
\bpm
0\\
1\\
0\\
-1 \\
\epm , \quad \,\,
v(\bos0, - \thalf) =
\frac{1}{\sqrt{2} }
\begin{pmatrix}
-1 \\ 0 \\ 1 \\ 0
\end{pmatrix} .
\label {the u, v spinors p=0}
\end{split}
\end{equation}
\par
They obey the parity conditions
\begin{equation}
u(\bos 0,s) ={} i \c^0 u(\bos 0,s) 
\qand
v(\bos 0,s) ={} - i \c^0 v(\bos 0,s) ,
\label {the parity conditions p=0}
\end{equation}
the charge-conjugation conditions
\begin{equation}
u(\bos 0,s) ={} \c^2 v^*(\bos 0,s) 
\qand
v(\bos 0,s) ={} \c^2 u^*(\bos 0,s) ,
\label {the Majorana conditions p=0}
\end{equation}
and the time-reversal conditions
\begin{equation}
u^*(\bos 0, s) = (-1)^{\shalf+ s}  
\c^1 \c^3 u(\bos 0, -s)
\qand
v^*(\bos 0, s) = (-1)^{\shalf+ s}  
\c^1 \c^3 v(\bos 0, -s) .
\label {the time-reversal conditions p=0}
\end{equation}
\par
The spinors (\ref{the u, v spinors p=0}) 
are derived 
in a book by Steven 
Weinberg~\citep{WeinbergIspinors} 
and in various articles~\citep{Dreiner:2008tw,PKCahill2006}, 
and they are discussed by 
Peskin and 
Schroeder~\citep{peskin1995AnIntroQFTspinors}
and by Srednicki~\citep{Srednicki2007spinors}\@.
But for reasons of space, style,  
or emphasis,
they are not stated explicitly in seven of the
leading textbooks on quantum field 
theory,
and they are
stated incorrectly in three of 
them~\citep{Ryder:1985wq,Kaku:1993ym,Schwartz:p189}\@.
\par
In those three books,
it is assumed that spinors merely
need to satisfy the Dirac equation
(\ref{the Dirac equation}).
But all that Dirac's equation says
about the spinors at zero
momentum is that the $u$ spinors
are eigenstates of $\c^0$
with eigenvalue $-i$,
and that the $v$ spinors
are eigenstates of $\c^0$
with eigenvalue $i$
\begin{equation}
\begin{split}
\c^0 \, u(\bos 0, s) ={}
-i \, u(\bos 0, s)
\qand
\c^0 \, v(\bos 0, s) ={}
i \, v(\bos 0, s) 
\end{split}
\end{equation}
which recapitulate
(\ref{equations for u, v spinors}--\ref{gamma0 and v spinors})\@.
But the eigenvalues $-i$ and $i$ 
are degenerate; each has two
eigenvectors.
The 
books~\citep{Ryder:1985wq,Kaku:1993ym,Schwartz:p189}  
interchanged the two eigenvectors
$v(\bos 0, \thalf)$
and $v(\bos 0, -\thalf)$\@.
\par
Using the wrong spinors leads 
to Dirac fields that don't transform
correctly under rotations,
Lorentz transformations,
charge conjugation, or
time reversal.
Such fields 
can lead to
physical results that are incorrect.
For instance, the naive 
spinors~\citep{Schwartz:p189}
\begin{equation}
 u_\uparrow ={}
 \frac{1}{\sqrt{2} }
\begin{pmatrix}
1 \\
0 \\
1 \\
0 \\
\end{pmatrix} , \quad
u_\downarrow ={}
\frac{1}{\sqrt{2} }
\bpm
0\\
1\\
0\\
1\\
\epm , \quad
v_\uparrow ={}
\frac{1}{\sqrt{2} }
\begin{pmatrix}
-1 \\ 0 \\ 1 \\ 0 
\end{pmatrix} ,
\qand
v_\downarrow ={}
\frac{1}{\sqrt{2} }
\begin{pmatrix}
0 \\ 1 \\ 0 \\ -1 
\end{pmatrix}
\label {naive spinors}
\end{equation}
don't obey the 
charge-conjugation conditions
(\ref{the Majorana conditions p=0})
and instead flip the spin
\begin{equation}
u_{\downarrow} ={}
\c^2 v^*_{\uparrow}, \quad
u_{\uparrow} ={}
\c^2 v^*_{\downarrow}, \quad
v_{\downarrow} ={}
\c^2 u^*_{\uparrow},
\qand
v_{\uparrow} ={}
\c^2 u^*_{\downarrow}.
\end{equation}
Their use 
leads to the false conclusion
that charge conjugation
(\ref{charge conjugation})
flips the spin~\citep{Schwartz:p189}.
The naive $v$ spinors 
(\ref{naive spinors})
also don't obey
the time-reversal conditions
(\ref{the time-reversal conditions p=0})
and instead introduce spurious signs:
$v_\uparrow = 
\c^1 \c^3 v_\downarrow$
and 
$v_\downarrow = 
- \c^1 \c^3 v_\uparrow$.
So Dirac and Majorana fields 
made with the naive 
spinors (\ref{naive spinors})
are mangled
under rotations,
Lorentz transformations,
charge conjugation,
and time reversal.

\section{Four-component spinors at finite momentum}
\label{Four-component spinors at finite momentum sec}

In this Section, 
we will use the 
zero-momentum,
4-component spinors
(\ref{the u, v spinors p=0})
and the
Dirac equation
in momentum space
to derive formulas for
the 4-component spinors
$u(\bos p,s)$ and $v(\bos p,s)$
at arbitrary momenta.
We will find that the $u$ 
and $v$ spinors differ more
at $\bos p \ne  \bos 0$ than 
at $\bos p = \bos 0$.
\par
The Dirac equation (\ref{the Dirac equation})
tells us how Dirac fields depend upon the coordinate
$x = (x^0, \bos x)$ and therefore how Dirac
spinors depend upon the momentum $p =(p^0, \bos p)$.
In particular, 
the combination 
\begin{equation}
\psi(x) = {} (m - \c^a \p_a) 
\, e^{\pm i p \cdot x} \chi 
= (m \mp i \c^a p_a) 
\, e^{\pm i p \cdot x} \chi
\label{chi}
\end{equation}
for $p^2 = -m^2$
obeys the Dirac equation 
(\ref{the Dirac equation})
for \textbf{\textit{every}} constant
four-component spinor 
$\chi$~\citep{CahillCUP2DiracEq2DiracEq}\@
\begin{equation}
\begin{split}
(\c^b \p_b + m) \psi(x) = {} &
(\pm i \c^b p_b + m) (m \mp i \c^a p_a) 
\, e^{\pm i p \cdot x} \chi 
= {}
( m^2 + p^2 ) \, e^{\pm i p \cdot x}
\chi 
= 0 .
\label{solution of Dirac equation}
\end{split}
\end{equation}
Since the $u$ spinors
in a Dirac field (\ref{Dirac 4-field})
occur with the phase $e^{ip\cdot x}$,
we set $\chi = u(\bos 0,s)$ and 
find that
\begin{equation}
\psi_{u, s}(x) 
= (m - \c^a \p_a) \, u(\bos 0,s) \, e^{ip\cdot x}
= (m - i \c^a p_a) \, u(\bos 0,s) \, e^{ip\cdot x}
\label{u(p,s)}
\end{equation}
is a solution of the Dirac equation
$(\c^b \p_b + m)\psi_{u, s}(x) = 0$.
So we set
\begin{equation}
u(\bos p, s) ={}
\frac{m - i \c^a p_a}{\sqrt{2p^0(p^0+m)}}
\, u(\bos 0,s)
\label {trick u spinor}
\end{equation}
in which the square root 
normalizes the spinor.
\par
Similarly since the $v$ spinors
in a Dirac field (\ref{Dirac 4-field})
occur with the phase $e^{-ip\cdot x}$,
we set 
$\chi = v(\bos 0,s)$ and find that
\begin{equation}
\psi_{v, s}(x) 
= (m - \c^a \p_a) \, v(\bos 0,s) \, e^{-ip\cdot x}
\label{v(p,s)}
\end{equation}
is a solution of the Dirac equation
$(\c^b \p_b + m)\psi_{v, s}(x) = 0$.
So we set
\begin{equation}
v(\bos p, s) ={} 
\frac{m + i \c^a p_a}{\sqrt{2p^0(p^0+m)}}
\, v(\bos 0,s) .
\label{trick v spinor}
\end{equation}
The vectors $u(\bos p, s)$ for $s = \pm \thalf$
are two eigenvectors of $-i \c^a p_a$ 
with eigenvalue $m$,
and the vectors $v(\bos p, s)$ for $s = \pm \thalf$
are two eigenvectors of $-i \c^a p_a$ 
with eigenvalue $-m$.
\par
In more detail,
the spinors 
are~\citep{PKCahill2006}
\begin{equation}
\begin{split}
u({\bos{p},\thalf}) ={}& 
\frac{1}{n(p^0)} \, 
\bpm
m+p^0-p_3 \\
         -p_1 - i p_2 \\
         m+p^0+p_3 \\
         p_1 +ip_2 
         \epm
         , \quad
u(\bos{p},-{\thalf}) ={} 
\frac{1}{n(p^0)} \, 
\bpm
-p_1+ip_2 \\
         m+p^0+p_3\cr
         p_1-ip_2\cr
         m+p^0-p_3\cr
         \epm
         \\
v({\bos{p},\thalf}) ={}& \frac{1}{n(p^0)} \, 
\bpm
-p_1+ip_2\cr
         m+p^0+p_3\cr
         -p_1+ip_2\cr
        p_3-m-p^0\cr
        \epm, \quad
v(\bos{p},{-\thalf}) = \frac{1}{n(p^0)} \, 
\bpm
p_3-m-p^0\cr
          p_1+ip_2\cr
          m+p^0+p_3\cr
          p_1+ip_2\cr
          \epm 
\label{u(p,s) v(p,s)} 
\end{split}
\end{equation}
in which the factor
$n(p^0) = 2\sqrt{p^0(p^0+m)}$
ensures their normalization
\begin{equation}
u^\dag(\bos p, s) \, u(\bos p, s') 
= {} \d_{s, s'},
\qquad
v^\dag(\bos p, s) \, v(\bos p, s') 
= {} \d_{s, s'}.
\label{They are normalized}
\end{equation}
\par
They obey 
the parity conditions
(\ref{the parity conditions p=0})
\begin{equation}
u(\bos p,s) ={} i \c^0 u(- \bos p,s) 
\qand
v(\bos p,s) ={} - i \c^0 v(- \bos p,s) ,
\label{the parity conditions p}
\end{equation}
the charge-conjugation conditions
(\ref{the Majorana conditions p=0})
\begin{equation}
u(\bos p,s) ={} \c^2 v^*(\bos p,s) 
\qand
v(\bos p,s) ={} \c^2 u^*(\bos p,s) ,
\label {the Majorana conditions p}
\end{equation}
and the time-reversal conditions
(\ref{the time-reversal conditions p=0})
\begin{equation}
u^*(\bos p, s) = (-1)^{\shalf+ s}  
\c^1 \c^3 u( - \bos p, -s)
\qand
v^*(\bos p, s) = (-1)^{\shalf+ s}  
\c^1 \c^3 v(- \bos p, -s) .
\label{the time-reversal conditions p}
\end{equation} 
\par
If one switches to a different set
of gamma matrices $\c'^i = S \, \c^i \, S^{-1}$,
then one must also switch one's spinors
to $u'(\bos p,s) = S \, u(\bos p,s)$ and 
$v'(\bos p,s)=S \, v(\bos p,s)$.

\section{Left-handed and right-handed spin-one-half fields}
\label{Left-handed and right-handed spin-one-half fields sec}

This Section describes how the upper
two components and the lower two components 
of a Dirac field transform as left-handed
and as right-handed fields
under Lorentz transformations.
\par
A 4-component
Dirac field $\psi_{\scriptscriptstyle D}(x)$,
$ {\scriptstyle D} = 1, 2, 3,4$,
\begin{equation}
\psi_{\scriptscriptstyle D}(x) 
={}
\sum_{s=\pm 1/2}
\int \frac{d^3p}{(2 \pi)^{3/2}} 
\lt[
u_{\scriptscriptstyle D}(\bos p, s) \, e^{ip\cdot x} \, a(\bos p,s)
+
v_{\scriptscriptstyle D}(\bos p, s) \, e^{-ip\cdot x} 
\, a_c^\dag(\bos p, s) \rt]
\label {4 component Dirac field}
\end{equation}
is made of a left-handed 2-component
field $\psi_\ell(x)$ and 
a right-handed 2-component
field $\psi_r(x)$
\begin{equation}
\psi(x) ={} 
\begin{pmatrix}
\psi_\ell(x) \\ \psi_r(x) 
\end{pmatrix}
\end{equation}
and so has the form
\begin{equation}
\begin{split}
\begin{pmatrix}
\psi_\ell(x) \\ \psi_r(x) 
\end{pmatrix} 
={}&
\sum_{s=\pm 1/2}
\int \frac{d^3p}{(2 \pi)^{3/2}} 
\lt[
\begin{pmatrix}
u_\ell(\bos p, s) \\
u_r(\bos p, s) 
\end{pmatrix}
e^{ip\cdot x} \, a(\bos p,s)
+
\begin{pmatrix}
v_\ell(\bos p, s) \\
v_r(\bos p, s) 
\end{pmatrix}
e^{-ip\cdot x} 
\, a_c^\dag(\bos p, s) \rt] .
\label {Dirac 4-field}
\end{split}
\end{equation}
\par
We can get a better understanding
of the left-handed and right-handed
spinors and fields by rewriting 
the spinor formulas
(\ref{trick u spinor} and
\ref{trick v spinor})\@.
To do that, we recall that
the zero-momentum spinors
(\ref{the u, v spinors p=0})
obey the parity conditions
(\ref{the parity conditions p=0})
\begin{equation}
u(\bos 0,s) ={} i \c^0 u(\bos 0,s) 
\qand
v(\bos 0,s) ={} - i \c^0 v(\bos 0,s) .
\label {the parity conditions p=0 redux}
\end{equation}
These conditions let us
replace $- i \c^a p_a \, u(\bos 0,s)$ 
by $\c^a \c^0 p_a \, u(\bos 0,s)$ and 
$ i \c^a p_a \, v(\bos 0,s)$
by $\c^a \c^0 p_a \, v(\bos 0,s)$ without
disturbing the definitions
(\ref{trick u spinor} and \ref{trick v spinor})
of the 4-spinors $u$ and $v$
\begin{equation}
u(\bos p, s) ={}
\frac{m + \c^a \c^0 p_a}{\sqrt{2p^0(p^0+m)}}
\, u(\bos 0,s)
\qand
v(\bos p, s) ={} 
\frac{m + \c^a \c^0 p_a}{\sqrt{2p^0(p^0+m)}}
\, v(\bos 0,s) .
\label{spinors at p in terms of those at 0}
\end{equation}
We now see that 
the 4-spinors $u(\bos p, s) $ and 
$v(\bos p, s)$ are 
generated from the
zero-momentum spinors
$u(\bos 0, s) $ and 
$v(\bos 0, s)$
by the same matrix 
$m + \c^a \c^0 p_a$
\begin{equation}
m + \c^a \c^0 p_a 
= {}
\begin{pmatrix}
m + p^0 - \bos p \cdot \bos \s & 0 \\
0 & m + p^0 + \bos p \cdot \bos \s
\end{pmatrix}
\end{equation}
which is block diagonal.
\par
The upper-left block 
$m + p^0 - \bos p \cdot \bos \s$
is proportional to the 
left-handed $2\by2$
representation 
$D^{(1/2,0)}(L(p))$ of the
Lorentz 
transformation $L(p)$
that takes momentum
$(m, \bos 0)$ to
$p = (p^0, \bos p)$
via a boost in the $\hat {\bos p}$
direction~\citep{WeinbergIspinors,PKCahill2006}.
The matrix 
$D^{(1/2,0)}(L(p))$ is
\begin{equation}
D^{(1/2,0)}(L(p))
={}
\frac{m + p^0 - \bos p \cdot \bos \s}
{\sqrt{2m(p^0+m)}}
=
\exp\lt( - \, \a \, \hat {\bos p} 
\cdot \frac{\bos \s}{2} \rt)
\end{equation}
in which 
$\a = \arctanh (|\bos p|/p^0)$~\citep{PKCahill2006,CahillCUP2DiracField}.
\par
The lower-right block 
$m + p^0 + \bos p \cdot \bos \s$
is proportional to the 
right-handed $2\by2$
representation 
$D^{(0,1/2)}(L(p))$ of the same
Lorentz transformation
$L(p)$ 
\begin{equation}
D^{(0,1/2)}(L(p))
={}
\frac{m + p^0 + \bos p \cdot \bos \s}
{\sqrt{2m(p^0+m)}}
=
\exp\lt( \, \a \, \hat {\bos p} 
\cdot \frac{\bos \s}{2} \rt).
\end{equation}
\par
If we combine the 
left- and right-handed
representations of $L(p)$
into a single $4\by4$ matrix
\begin{equation}
D^{(1/2, 0) \oplus (0, 1/2)}(L(p))
={}
\begin{pmatrix}
D^{(1/2,0)}(L(p)) & 0 \\
0 & D^{(0,1/2)}(L(p)) 
\end{pmatrix},
\label {4x4 matrix L(p)}
\end{equation}
then we can write the formulas
(\ref{spinors at p in terms of those at 0})
for the spinors as
\begin{equation}
\begin{split}
u(\bos p, s) ={} &
\sqrt{\frac{m}{p^0}} \,
D^{(1/2, 0) \oplus (0, 1/2)}(L(p)) \, 
u(\bos 0, s) 
\\
v(\bos p, s) ={} &
\sqrt{\frac{m}{p^0}} \,
D^{(1/2, 0) \oplus (0, 1/2)}(L(p)) \, 
v(\bos 0, s) 
\label {u,v = D u,v0}
\end{split}
\end{equation}
or equivalently as
\begin{equation}
\begin{split}
u(\bos p, s) ={} &
\begin{pmatrix}
u_\ell(\bos p, s) \\ u_r(\bos p, s)
\end{pmatrix}
=
\sqrt{\frac{m}{p^0}} \,
\begin{pmatrix}
D^{(1/2,0)}(L(p)) \, u_\ell(\bos 0, s) \\
D^{(0,1/2)}(L(p)) \, u_r(\bos 0, s)
\end{pmatrix}
\\
v(\bos p, s) ={} &
\begin{pmatrix}
v_\ell(\bos p, s) \\ v_r(\bos p, s)
\end{pmatrix}
=
\sqrt{\frac{m}{p^0}} \,
\begin{pmatrix}
D^{(1/2,0)}(L(p)) \, v_\ell(\bos 0, s) \\
D^{(0,1/2)}(L(p)) \, v_r(\bos 0, s)
\end{pmatrix}.
\label{left and right spinors}
\end{split}
\end{equation}

\par
Weinberg~\citep{WeinbergIspinorslong} 
has shown that 
a Dirac field (\ref{Dirac 4-field})
transforms  under 
a Lorentz transformation $\La$
as
\begin{equation}
U(\La) \, \psi_{\scriptscriptstyle{D}}(x) 
\, U^{-1}(\La)
={}
D^{(1/2, 0) \oplus (0, 1/2)}
_{\scriptscriptstyle{D}, \scriptscriptstyle{D}'}
(\La^{-1}) \, 
\sum_{\scriptscriptstyle{D}' =1}^4
\psi_{\scriptscriptstyle{D}'}(\La x) 
\label {Lorentz transformation}
\end{equation}
if and only if 
the spinors are defined
by equations 
(\ref{the u, v spinors p=0} and
\ref{u,v = D u,v0}).
A derivation of this result
is sketched in 
appendix~\ref{Wigner rotations sec}.
\par
Since the $4\by4$ matrix
$D^{(1/2, 0) \oplus (0, 1/2)}
_{\scriptscriptstyle{D}, \scriptscriptstyle{D}'}
(\La^{-1})$ is block diagonal,
we can separate equation
(\ref{Lorentz transformation}),
which says how a Dirac field
transforms under a Lorentz
transformation $\La$,
into one equation for the 
left-handed field $\psi_\ell$
and another for the 
right-handed field $\psi_r$
\begin{equation}
\begin{split}
U(\La) \psi_\ell(x) 
U^{-1}(\La)
={}&
D^{(1/2,0)}(\La^{-1}) \,
\psi_\ell(\La x)
\\
U(\La) \psi_r(x) 
U^{-1}(\La)
={}&
D^{(0,1/2)}(\La^{-1}) \,
\psi_r(\La x) .
\end{split}
\end{equation}
\par
To see why $u_\ell$ and $v_\ell$ 
are called left handed 
and why $u_r$ and $v_r$ 
are called right handed, 
we look at the terms 
$u_\ell(\bos p, s) 
\, e^{ip\cdot x} \, a(\bos p,s)$,
$v_\ell(\bos p, s) \, e^{-ip\cdot x} 
\, a_c^\dag(\bos p, s)$,
$u_r(\bos p, s) 
\, e^{ip\cdot x} \, a(\bos p,s)$,
and
$v_r(\bos p, s) \, e^{-ip\cdot x} 
\, a_c^\dag(\bos p, s)$
in the Dirac field
(\ref{Dirac 4-field})
for particles with momentum 
in the $z$ direction
$\bos p = p \hat {\bos z}$
in the limit $m/p^0 \to 0$,
a limit reached by neutrinos
with $p^0 \gtrsim 1$ keV\@.
In the $m/p^0 \to 0$ limit, the spinors
(\ref{u(p,s) v(p,s)}) for
$\bos p = p \hat {\bos z}$
are
\begin{equation}
\begin{split}
\begin{pmatrix}
u_\ell(p \hat {\bos z},\thalf) \\
u_r(p \hat {\bos z},\thalf)
\end{pmatrix} 
\to {}&
\begin{pmatrix}
0\cr 
         0\cr
         1 \cr 
         0\\ \end{pmatrix}
, \qquad
\begin{pmatrix}
u_\ell(p \hat {\bos z}, -\thalf) \\
u_r(p \hat {\bos z}, -\thalf)
\end{pmatrix} 
\to 
\begin{pmatrix}
0\cr 
         1 \cr
         0\cr 
         0\\ \end{pmatrix}
         , 
         \\   
\begin{pmatrix}
v_\ell(p \hat {\bos z},\thalf) \\
v_r(p \hat {\bos z},\thalf)
\end{pmatrix}            
\to {}&
\begin{pmatrix}
0\cr 
         1 \cr
         0\cr 
         0\\ \end{pmatrix}
         , \qand
         \begin{pmatrix}
v_\ell(p \hat {\bos z}, -\thalf) \\
v_r(p \hat {\bos z}, -\thalf)
\end{pmatrix}   
\to 
\begin{pmatrix}
0\cr 
          0\cr
          1 \cr 
          0\\ \end{pmatrix}
\label{u, v spinors in small m limit}
\end{split}
\end{equation}
and so the nonzero terms 
$u_{\scriptscriptstyle D}(\bos p, s) \, e^{ip\cdot x} \, a(\bos p,s)$
and
$v_{\scriptscriptstyle D}(\bos p, s) \, e^{-ip\cdot x} 
\, a_c^\dag(\bos p, s)$
are
\begin{equation}
\begin{split}
u_r(p \hat {\bos z},\thalf) &
\, e^{ip \cdot x} \, a(p \hat {\bos z},\thalf),
\quad
u_\ell(p \hat {\bos z}, -\thalf)
\, e^{ip \cdot x} \, a(p \hat {\bos z}, -\thalf),
\\
v_\ell(p \hat {\bos z},\thalf) &
\, e^{-ip\cdot x} 
\, a_c^\dag(p \hat {\bos z}, \thalf)
, \qand
v_r(p \hat {\bos z}, -\thalf)
\, e^{-ip\cdot x} 
\, a_c^\dag(p \hat {\bos z}, -\thalf).
\label{u,v in big p limit}
\end{split}
\end{equation}
The field $\ell(x)$ is said to be left handed
because it
contains $u_\ell(p \hat {\bos z}, -\thalf)
\, a(p \hat {\bos z}, -\thalf)$ 
which destroys particles
with momentum 
antiparallel to the spin
and $v_\ell(p \hat {\bos z},\thalf) 
\, a_c^\dag(p \hat {\bos z}, \thalf)$
which creates antiparticles
with momentum 
parallel to the spin.
The field $r(x)$ is said to be right handed
because it contains
$u_r(p \hat {\bos z},\thalf) 
\, a(p \hat {\bos z},\thalf)$
which destroys particles
with momentum 
parallel to the spin
and $v_r(p \hat {\bos z}, -\thalf)
\, a_c^\dag(p \hat {\bos z}, -\thalf)$
which creates antiparticles
with momentum 
antiparallel to the spin.
The weak gauge group
$SU(2)_\ell$ acts on 
left-handed fields.
\par
The simplest spin-one-half 
fields are the 2-component
Majorana fields that are 
linear combinations
of the annihilation
operators $a(\bos p,s)$
and their adjoints
$a^\dag(\bos p,s)$
multiplied by left-handed
and right-handed 
2-component spinors
(\ref{left and right spinors})
\begin{equation}
\begin{split}
\psi_{\ell \scriptscriptstyle{M}}(x) 
= {}&
\sum_{s = \pm 1/2} 
\int \frac{d^3p}{(2 \pi)^{3/2}} \>\;
u_\ell(\bos p, s) \, e^{ip\cdot x} \, a(\bos p,s)
+
v_\ell(\bos p, s) \, 
e^{-ip\cdot x} \, a^\dag(\bos p, s)
\\
\psi_{r \scriptscriptstyle{M}}(x) 
= {}&
\sum_{s = \pm 1/2} 
\int \frac{d^3p}{(2 \pi)^{3/2}} \>\;
u_r(\bos p, s) \, e^{ip\cdot x} \, a(\bos p,s)
+
v_r(\bos p, s) \, 
e^{-ip\cdot x} \, a^\dag(\bos p, s) .
\label{2-component Majorana fields}
\end{split}
\end{equation}
Together they make
a 4-component Majorana field
\begin{equation}
\psi_{\scriptscriptstyle{M}}(x) 
={}
\begin{pmatrix}
\psi_{\ell \scriptscriptstyle{M}}(x) 
\\
\psi_{r \scriptscriptstyle{M}}(x)  
\end{pmatrix} .
\label{4-component Majorana field}
\end{equation}
\par
If one has two annihilation operators
$a_1(\bos p, s)$ and $a_2(\bos p, s)$
and their adjoints
$a_1^\dag(\bos p, s)$ and 
$a_2^\dag(\bos p, s)$,
all referring to 
particles of the same mass, 
then one may make operators
that annihilation and create
particles and their antiparticles
\begin{equation}
\begin{split}
a(\bos p,s) ={}& \frac{1}{\sqrt{2}}
\Big[ a_1(\bos p,s) + i a_2(\bos p,s) \Big],
\quad
a_c(\bos p,s) ={} \frac{1}{\sqrt{2}}
\Big[ a_1(\bos p,s) - i a_2(\bos p,s) \Big]
\\
a^\dag(\bos p,s) ={}& \frac{1}{\sqrt{2}}
\Big[ a_1^\dag\bos p,s) - i a_2^\dag(\bos p,s) 
\Big],
\quad
a_c^\dag(\bos p,s) ={} \frac{1}{\sqrt{2}}
\Big[ a_1^\dag(\bos p,s) + i a_2^\dag(\bos p,s) 
\Big] 
\end{split}
\end{equation}
and define a Dirac field as
\begin{equation}
\psi(x) ={} \frac{1}{\sqrt{2}}
\lt[ \psi_{\scriptscriptstyle{M} 1}(x) 
+ i \psi_{\scriptscriptstyle{M} 1}(x)
\rt]
=  \begin{pmatrix}
\psi_\ell(x)
\\
\psi_r(x)
\end{pmatrix}
= \frac{1}{\sqrt{2}}
\begin{pmatrix}
\psi_{\ell \scriptscriptstyle{M} 1}(x) 
+ i \psi_{\ell \scriptscriptstyle{M} 2}(x) 
\\
\psi_{r \scriptscriptstyle{M} 1}(x) 
+ i \psi_{r \scriptscriptstyle{M} 1}(x) 
\end{pmatrix}  .
\label{Dirac field as 2 Majorana fields}
\end{equation}

\section{Conclusions}
\label{Conclusions sec}

The Dirac spinors at momentum zero
$u(\bos 0, s)$ and $v(\bos 0, s)$ 
cannot be chosen arbitrarily as four 
orthonormal 4-component vectors.
They are instead 
determined 
(\ref{the u, v spinors p=0})
by the requirement that
the Dirac field transform 
correctly under rotations
and obey the Dirac equation
as explained in 
Sections~\ref{Rotations of states of zero momentum and spin one-half sec}--\ref{Four-component spinors at zero momentum sec}\@. 
\par
Once one has the 
zero-momentum spinors, 
the Dirac equation yields
the finite-momentum spinors 
(\ref{trick u spinor})
and (\ref{trick v spinor})
as explained in Section~\ref{Four-component spinors at finite momentum sec}\@.
\par
A 4-component
Dirac field $\psi$ is composed of a 
2-component left-handed
field $\psi_\ell$ and a 
2-component right-handed
field $\psi_r$
as described in 
Section ~\ref{Left-handed and right-handed spin-one-half fields sec}\@.
\par
Appendix~\ref{Charge conjugation, parity, and time reversal sec} shows that when spinors are defined
correctly (\ref{the u, v spinors p=0}, \ref{trick u spinor}, and 
\ref{trick v spinor}), a Dirac field transforms appropriately 
under charge conjugation, parity, and time reversal.
Appendix~\ref{Wigner rotations sec} explains
how particles and fields transform under
Lorentz transformations and shows that the
spinors of a Dirac field that transforms correctly
under Lorentz transformations has spinors
that are related to its zero-momentum
spinors by equations (\ref{trick u spinor})
and (\ref{trick v spinor})\@.
Appendix~\ref{Majorana and Dirac Fields sec}
explains that a Dirac field is a complex linear
combination of two Majorana fields of the 
same mass.

\begin{acknowledgements}
I'd like to thank 
Rouzbeh Allahverdi for
helpful email and
the referees 
for their suggestions
which improved this paper.
\end{acknowledgements}

\appendix

\section{Charge conjugation, parity, and time reversal}
\label{Charge conjugation, parity, and time reversal sec}

Because spinors obey the parity,
charge-conjugation, and time-reversal 
conditions
(\ref{the parity conditions p}, 
\ref{the Majorana conditions p}, and 
\ref{the time-reversal conditions p}),
Dirac fields transform simply
under parity,
charge conjugation, and time reversal.
\par
Parity reverses space and 
momentum,
so the annihilation and creation
operators obey the 
rules~\citep{WeinbergIparity}
\begin{equation}
\mathsf{P} \, a_c^\dag(\bos p,s) 
\, \mathsf{P}^{-1} ={}
\eta_c \, a_c^\dag(- \bos p,s) 
= - \eta^* \, a_c^\dag(- \bos p,s)
\qand
\mathsf{P} \, a(\bos p,s) 
\, \mathsf{P}^{-1} ={}
\eta^* \, a(- \bos p,s) .
\label {parity}
\end{equation}
They and
the parity conditions
(\ref{the parity conditions p})
imply that a Dirac field transforms
simply under parity
\begin{align}
\mathsf{P} \, \psi(t, \bos x) 
\, \mathsf{P}^{-1} 
={}& 
\sum_s \int \frac{d^3p}{(2 \pi)^{3/2}} \>
u(\bos p, s) \, e^{ip\cdot x} \, 
\mathsf{P} a(\bos p,s)
\mathsf{P}^{-1} 
+
v(\bos p, s) \, e^{-ip\cdot x} \, 
\mathsf{P} a_c^\dag(\bos p, s) 
\mathsf{P}^{-1} 
\nn\\
={}& 
\sum_s \int \frac{d^3p}{(2 \pi)^{3/2}} \>
u(\bos p, s) \, e^{ip\cdot x} \, 
\eta^* a(- \bos p,s)
+
v(\bos p, s) \, e^{-ip\cdot x} \, 
\eta_c \, a_c^\dag(- \bos p, s) 
\nn\\
={}& 
\sum_s \int \frac{d^3p}{(2 \pi)^{3/2}} \>
u(- \bos p, s) \, e^{ip\cdot Px} \, 
\eta^* a(\bos p,s)
+
v(- \bos p, s) \, e^{-ip\cdot Px} \, 
\eta_c \, a_c^\dag(\bos p, s) 
\nn\\
={}& 
\sum_s \int \frac{d^3p}{(2 \pi)^{3/2}} \>
i \c^0 u(\bos p, s) \, e^{ip\cdot Px} \, 
\eta^* a(\bos p,s)
-
i \c^0 v(\bos p, s) \, e^{-ip\cdot Px} \, 
\eta_c \, a_c^\dag(\bos p, s) 
\nn\\
={}& 
\eta^* i \c^0
\sum_s \int \frac{d^3p}{(2 \pi)^{3/2}} \>
u(\bos p, s) \, e^{ip\cdot Px} \, 
a(\bos p,s)
+
v(\bos p, s) \, e^{-ip\cdot Px} \, 
a_c^\dag(\bos p, s) 
\nn\\
={}& 
\eta^* i \c^0 \, \psi(t, - \bos x) .
\end{align}
\par
A Majorana field (\ref{4-component Majorana field})
creates and destroys
the same kind of particle.
Its Fourier expansion is 
that of a Dirac field but
with $a_c^\dag = a^\dag$
\begin{equation}
   \begin{split}
\psi_{\scriptscriptstyle{M}}(x) = {}&
\sum_{s = \pm 1/2} 
\int \frac{d^3p}{(2 \pi)^{3/2}} \>\;
u(\bos p, s) \, e^{ip\cdot x} \, a(\bos p,s)
+
v(\bos p, s) \, 
e^{-ip\cdot x} \, a^\dag(\bos p, s) .
\label {A Majorana field}
   \end{split}
\end{equation}
For a Majorana particle,
$a_c^\dag(\bos p,s) = a^\dag(\bos p,s)$,
so $\eta_c = \eta$\@.
But the parity rules (\ref{parity})
also require that
$\eta_c = - \eta^*$, 
so $\eta$ 
must be imaginary,
$\eta= - \eta^*$\@.
\par
In general, charge conjugation maps
particles into their 
antiparticles~\citep{WeinbergIC}
\begin{equation}
\mathsf{C} \, a_c^\dag(p, s) \, \mathsf{C}^{-1} 
= {} \xi_c a^\dag(p, s) 
= {} \xi^* a^\dag(p, s) 
\qand
\mathsf{C} \, a(p, s) \, \mathsf{C}^{-1} = {} 
\xi^* a_c(p, s) .
\label {charge conjugation}
\end{equation}
The charge-conjugation conditions
(\ref{the Majorana conditions p})
and the definition 
(\ref{charge conjugation})
of charge conjugation
imply that a Dirac field
(\ref{4 component Dirac field}) 
transforms simply 
under charge conjugation
\begin{equation}
   \begin{split}
\mathsf{C} \, \psi(x) \, \mathsf{C}^{-1} 
= {}&
\sum_s \int \frac{d^3p}{(2 \pi)^{3/2}} \>
u(\bos p, s) \, e^{ip\cdot x} \, 
\mathsf{C} a(\bos p,s)
\mathsf{C}^{-1} 
+
v(\bos p, s) \, e^{-ip\cdot x} \, 
\mathsf{C} a_c^\dag(\bos p, s) 
\mathsf{C}^{-1} 
\\
= {}&
\sum_s \int \frac{d^3p}{(2 \pi)^{3/2}} \>
u(\bos p, s) \, e^{ip\cdot x} \, 
\xi^* a_c(\bos p,s)
+
v(\bos p, s) \, e^{-ip\cdot x} \, 
\xi_c a^\dag(\bos p, s) 
\\
= {}&
\xi^* \sum_s \int \frac{d^3p}{(2 \pi)^{3/2}} \>
\c^2 v^*(\bos p,s) \, e^{ip\cdot x} \, 
a_c(\bos p,s)
+
\c^2 u^*(\bos p,s) \, e^{-ip\cdot x} \, 
a^\dag(\bos p, s) 
\\
= {}&
\xi^* \c^2  \sum_s \int \frac{d^3p}
{(2 \pi)^{3/2}} \>\;
u^*(\bos p, s) \, 
e^{-ip\cdot x} \, a^\dag(\bos p)
+
v^*(\bos p, s) \, 
e^{ip\cdot x} \, a_c(\bos p, s) 
\\
= {}&
\xi^* \c^2 \, \psi^*(x) 
\label {massive Dirac field and charge conjugation}
   \end{split}
\end{equation}
in which the asterisk denotes
complex and hermitian conjugation
but not the conversion of column vectors
into row vectors.
\par
For Majorana particles, 
the charge-conjugation conditions
(\ref{the Majorana conditions p})
imply that
$\xi_c a^\dag(p, s) = 
\mathsf{C} \, a_c^\dag(p, s) \, \mathsf{C}^{-1}
= \mathsf{C} a^\dag(p, s) 
\mathsf{C}^{-1} = \xi a^\dag(p, s)$,
so the phase
$\xi_c = \xi^* = \xi$
is real for Majorana particles.
The charge-conjugation conditions
(\ref{the Majorana conditions p})
also imply that hermitian conjugation
changes a Majorana field 
(\ref{A Majorana field}) to
\begin{equation}
\begin{split}
\psi^*_{\scriptscriptstyle M}
={}&
\sum_{s = \pm 1/2} 
\int \frac{d^3p}{(2 \pi)^{3/2}} \>\;
u^*(\bos p, s) \, e^{-ip\cdot x} 
\, a^\dag(\bos p,s)
+
v^*(\bos p, s) \, 
e^{ip\cdot x} \, a(\bos p, s)
\\
={}&
\sum_{s = \pm 1/2} 
\int \frac{d^3p}{(2 \pi)^{3/2}} \>\;
\c^2 
u(\bos p, s) \, e^{ip\cdot x} \, a(\bos p,s)
+
\c^2 v(\bos p, s) \, 
e^{-ip\cdot x} \, a^\dag(\bos p, s) 
\end{split}
\end{equation}
so that a Majorana field obeys
the Majorana condition
\begin{equation}
\psi^*_{\scriptscriptstyle M}(x) 
= {} \c^2 \, \psi_{\scriptscriptstyle M}(x) .
\label{obeys the Majorana condition}
\end{equation}

\par
Time reversal reverses momentum
and spin, adds a phase, and complex
conjugates complex 
numbers~\citep{WeinbergICPT}
\begin{equation}
\begin{split}
\mathsf{T} \, z \, a(\bos p, s) \, \mathsf{T}^{-1} 
={}& 
z^* \, 
(-1)^{\shalf -s} \, \zeta^* \, 
a(- \bos p, -s)
\\
\mathsf{T} \, w \, a_c^\dag(\bos p, s) \, \mathsf{T}^{-1} 
={}& 
w^* \, 
(-1)^{\shalf -s} \, \zeta_c 
\, a_c^\dag(- \bos p, -s) 
\label{time-reversal}
\end{split}
\end{equation}
in which $z$ and $w$ are arbitrary
complex numbers, and
$\zeta_c = \zeta^*$.
This definition 
(\ref{time-reversal}) and
the time-reversal conditions
(\ref{the time-reversal conditions p})
imply that a Dirac field
(\ref{4 component Dirac field}) 
transforms simply 
under time reversal
\begin{align}
\mathsf{T} \, \psi(t,\bos x) \, & \mathsf{T}^{-1} 
= {} 
\sum_s \int \frac{d^3p}{(2 \pi)^{3/2}} \>
u^*(\bos p, s) \, e^{-ip\cdot x} \, 
\mathsf{T} a(\bos p,s)
\mathsf{T}^{-1} 
+
v^*(\bos p, s) \, e^{ip\cdot x} \, 
\mathsf{T} a_c^\dag(\bos p, s) 
\mathsf{T}^{-1} 
\nn \\
= {}& 
\sum_s 
(-1)^{\shalf -s} \!\!
\int \!
\frac{d^3p}{(2 \pi)^{3/2}} \,
u^*(\bos p, s) \, e^{-ip\cdot x} 
\zeta^* \, a(- \bos p, -s)
+
v^*(\bos p, s) \, e^{ip\cdot x}
\zeta_c \, a_c^\dag(- \bos p, -s) 
\nn \\
= {}& 
\sum_s 
(-1)^{\shalf -s} \!\!
\int \!
\frac{d^3p}{(2 \pi)^{3/2}} \,
u^*(\bos p, s) \, e^{-ip\cdot x} 
\zeta^* \, a(- \bos p, -s)
+
v^*(\bos p, s) \, e^{ip\cdot x}
\zeta^* \, a_c^\dag(- \bos p, -s) \nn \\
= {}& 
\zeta^*
\sum_s 
(-1)^{\shalf -s} \!\!
\int \!
\frac{d^3p}{(2 \pi)^{3/2}} \,
u^*(\bos p, s) \, e^{-ip\cdot x} 
a(- \bos p, -s)
+
v^*(\bos p, s) \, e^{ip\cdot x}
a_c^\dag(- \bos p, -s) \nn\\
= {}& 
\zeta^* \c^1 \c^3
\sum_s \! 
\int \!
\frac{d^3p}{(2 \pi)^{3/2}} \,
u(-\bos p, -s) \, e^{-ip\cdot x} 
a(- \bos p, -s)
+
v(- \bos p, -s) \, e^{ip\cdot x}
a_c^\dag(- \bos p, -s) 
\nn\\
= {}& 
\zeta^* \c^1 \c^3 \,
\psi(- t, \bos x) .
\label {massive Dirac field and time reversal}
\end{align}

\section{Wigner rotations}
\label{Wigner rotations sec}

Steven Weinberg has shown that the
Lorentz-transformation properties
of a quantum field of any spin 
determine the spinors of that 
field~\citep{WeinbergIderivation}\@.
This appendix repeats that 
derivation for fields of 
spin one-half and inserts
some extra steps that 
may help students.
\par
In 
Section~\ref{Rotations of states of zero momentum and spin one-half sec} 
we considered how rotations
change states $| \bos 0, s \ra$
without saying how Lorentz boosts
change them to states of finite momentum.  
Because we'll be 
talking about Lorentz transformations,
it will be convenient to write
states as
$| (p^0,\bos p), s \ra \equiv |p, s \ra$\@.
The state $| p, s \ra$
is the image of the state 
of a particle at rest $| 0, s \ra$
under the 
unitary transformation $U(L(p))$
that implements the standard
Lorentz boost $L(p)$ in the
direction $\bos p$
\begin{equation}
| p, s \ra ={}
\sqrt{\frac{m}{p^0}} \, 
U(L(p)) \, | 0, s \ra
\label{is the image of the state}
\end{equation}
normalized so that 
$\la p, s | p', s' \ra =
\d_{s s'} \d({\bos p} - {\bos p}')$\@.
A Lorentz transformation $\La$
changes the state $| p, s \ra$
to a linear combination
of states with momentum
$\La p$ but with 
different spins in the $z$ direction 
\begin{equation}
U(\La) | p, s \ra 
={}
\sqrt{\frac{(\La p)^0}{p^0}}
\sum_{s' = \pm 1/2} 
D_{s' s}(R_{\scriptscriptstyle W}(\La, p))
\, | \La p, s' \ra .
\end{equation}
The $2\by2$ matrix 
$D(R_{\scriptscriptstyle W}(\La, p))$ 
as in equation~(\ref{matrix D_z})
represents the Wigner 
rotation~\citep{WignerRotation,WeinbergIrelativisticQM} 
\begin{equation}
R_{\scriptscriptstyle W}(\La, p)
={}
L^{-1}(\La p) \, \La \, L(p)
\end{equation}
that boosts a particle
at rest to momentum $p$,
and then to $\La p$,
and then back to rest.
The Wigner rotation arises because
\begin{equation}
\begin{split}
U(\La) | p, s \ra 
={}&
N(p) \, U(\La p)
\, U(L^{-1}(\La p)) 
\, U(\La) \, U(L(p)) \, | 0, s \ra
\\
={}&
N(p) \, U(\La p)
\, U(R_{\scriptscriptstyle W}(\La, p)) \, | 0, s \ra
\\
={}&
N(p) \, U(\La p)
\sum_{s' = -1/2}^{1/2} 
D_{s' s}(R_{\scriptscriptstyle W}(\La, p))
\, | 0, s' \ra
\\
={}&
\sqrt{\frac{(\La p)^0}{p^0}}
\sum_{s' = -1/2}^{1/2} 
D_{s' s}(R_{\scriptscriptstyle W}(\La, p))
\, | \La p, s' \ra.
\end{split}
\end{equation}
So the generalizations
to Lorentz transformations
of states of finite momentum
of the formulas
(\ref{how a and adag go under rotations})
for how zero-momentum
creation and annihilation operators
transform under rotations are
\begin{equation}
\begin{split}
U(\La) \, a^\dag(p,s)
\, U^{-1}(\La)
={}&
\sqrt{\frac{(\La p)^0}{p^0}}
\sum_{s' = -1/2}^{1/2} 
D^*_{s s'}(R^{-1}_{\scriptscriptstyle W}(\La, p))
\, a^\dag(\La p,s)
\\
U(\La) \, a(p,s)
\, U^{-1}(\La)
={}&
\sqrt{\frac{(\La p)^0}{p^0}}
\sum_{s' = -1/2}^{1/2} 
D_{s s'}(R^{-1}_{\scriptscriptstyle W}(\La, p))
\, a(\La p, s').
\label{L on a a*}
\end{split}
\end{equation}
Thus a Dirac field 
(\ref{4 component Dirac field})
will transform correctly
under a Lorentz
transformation $\La$
(\ref{Lorentz transformation})
\begin{equation}
\begin{split}
U(\La) \, \psi_{\scriptscriptstyle D}(x)
\, U^{-1}(\La)
={}&
\sum_{s=\pm 1/2}
\int \frac{d^3p}{(2 \pi)^{3/2}} 
\Big[
u_{\scriptscriptstyle D}(p, s) \, e^{ip\cdot x} \, U(\La) a(p,s) U^{-1}(\La)
\\
{}&+
v_{\scriptscriptstyle D}(p, s) \, e^{-ip\cdot x} 
\, U(\La) a_c^\dag(p, s) 
U^{-1}(\La)\Big]
\\
={}&
\sum_{\scriptscriptstyle{D}' =1}^4
D^{(1/2, 0) \oplus (0, 1/2)}
_{\scriptscriptstyle{D}, \scriptscriptstyle{D}'}
(\La^{-1})
\psi_{\scriptscriptstyle{D}'}(\La x) 
\\
={}&
\sum_{{\scriptscriptstyle{D}'}=1}^4
\sum_{s=\pm 1/2}
D^{(1/2, 0) \oplus (0, 1/2)}
_{\scriptscriptstyle{D}, \scriptscriptstyle{D}'}
(\La^{-1})
\\
{}& \by
\int \frac{d^3p}{(2 \pi)^{3/2}} 
\lt[
u_{\scriptscriptstyle D'}(p, s) \, e^{ip\cdot \La x} \, a(p,s)
+
v_{\scriptscriptstyle D'}(p, s) \, e^{-ip\cdot \La x} 
\, a_c^\dag(p, s) \rt]
\end{split}
\end{equation}
if 
\begin{equation}
\begin{split}
\sum_{s, s'=\pm 1/2}
\int & \frac{d^3p}{(2 \pi)^{3/2}} 
\sqrt{\frac{(\La p)^0}{p^0}}
u_{\scriptscriptstyle D}(p, s) \, e^{ip\cdot x}
D_{s s'}(R^{-1}_{\scriptscriptstyle W}(\La, p))
\, a(\La p, s')
\\
={}&
\sum_{{\scriptscriptstyle{D}'}=1}^4
\sum_{s=\pm 1/2}
D^{(1/2, 0) \oplus (0, 1/2)}
_{\scriptscriptstyle{D}, \scriptscriptstyle{D}'}
(\La^{-1})
\int \frac{d^3p}{(2 \pi)^{3/2}} 
u_{\scriptscriptstyle D'}(p, s) 
\, e^{ip\cdot \La x} \, a(p,s)
\label{try a}
\end{split}
\end{equation}
and
\begin{equation}
\begin{split}
\sum_{s, s'=\pm 1/2}
\int & \frac{d^3p}{(2 \pi)^{3/2}} 
\sqrt{\frac{(\La p)^0}{p^0}}
v_{\scriptscriptstyle D}(p, s) \, e^{-ip\cdot x}
D^*_{s s'}(R^{-1}_{\scriptscriptstyle W}(\La, p))
\, a^\dag_c(\La p, s')
\\
={}&
\sum_{{\scriptscriptstyle{D}'}=1}^4
\sum_{s=\pm 1/2}
D^{(1/2, 0) \oplus (0, 1/2)}
_{\scriptscriptstyle{D}, \scriptscriptstyle{D}'}
(\La^{-1})
\int \frac{d^3p}{(2 \pi)^{3/2}} 
v_{\scriptscriptstyle D'}(p, s) \, e^{-ip\cdot \La x} \, a^\dag_c(p,s).
\label{try a*}
\end{split}
\end{equation}
Setting $ d^3 p = p^0 d^3 \La p /(\La p)^0)$
in the left-hand sides
of these equations 
(\ref{try a} and \ref{try a*}),
and then on their 
right-hand sides changing variables
$p \to \La p$ and using
$\La p \cdot \La x = p \cdot x$,
we get
\begin{equation}
\begin{split}
\sum_{s, s'=\pm 1/2}
\int & \frac{d^3\La p}{(2 \pi)^{3/2}} 
\sqrt{\frac{p^0}{(\La p)^0}}
u_{\scriptscriptstyle D}(p, s) \, e^{ip\cdot x}
D_{s s'}(R^{-1}_{\scriptscriptstyle W}(\La, p))
\, a(\La p, s')
\\
={}&
\sum_{{\scriptscriptstyle{D}'}=1}^4
\sum_{s=\pm 1/2}
D^{(1/2, 0) \oplus (0, 1/2)}
_{\scriptscriptstyle{D}, \scriptscriptstyle{D}'}
(\La^{-1})
\int \frac{d^3 \La p}{(2 \pi)^{3/2}} 
\, u_{\scriptscriptstyle D'}(\La p, s) 
\, e^{ip\cdot x} \, a(\La p,s)
\label{try a redux}
\end{split}
\end{equation}
and
\begin{equation}
\begin{split}
\sum_{s, s'=\pm 1/2}
\int & \frac{d^3 \La p}{(2 \pi)^{3/2}} 
\sqrt{\frac{p^0}{(\La p)^0}}
\, v_{\scriptscriptstyle D}(p, s) \, e^{-ip\cdot x}
D^*_{s s'}(R^{-1}_{\scriptscriptstyle W}(\La, p))
\, a^\dag_c(\La p, s')
\\
={}&
\sum_{{\scriptscriptstyle{D}'}=1}^4
\sum_{s=\pm 1/2}
D^{(1/2, 0) \oplus (0, 1/2)}
_{\scriptscriptstyle{D}, \scriptscriptstyle{D}'}
(\La^{-1})
\int \frac{d^3 \La p}{(2 \pi)^{3/2}} 
\, v_{\scriptscriptstyle D'}(\La p, s) 
\, e^{-ip\cdot x} \, a^\dag_c(\La p,s).
\label{try a* redux}
\end{split}
\end{equation}
By equating the coefficients
of $e^{ip\cdot x} a(\La p,s)$
in (\ref{try a redux})
and of 
$e^{-ip\cdot x} a^\dag_c(\La p,s)$
in (\ref{try a* redux}), we find
\begin{equation}
\begin{split}
\sum_{s'} 
\sqrt{\frac{p^0}{(\La p)^0}} 
u_{\scriptscriptstyle D}(p, s') \, 
D^{-1}_{s' s}(R_{\scriptscriptstyle W}(\La, p))
={}&
\sum_{\scriptscriptstyle{D'}}
D^{(1/2, 0) \oplus (0, 1/2) -1}
_{\scriptscriptstyle{D}, \scriptscriptstyle{D}'}
(\La)
\, u_{\scriptscriptstyle D'}(\La p, s) 
\\
\sum_{s'}
\sqrt{\frac{p^0}{(\La p)^0}}
\, v_{\scriptscriptstyle D}(p, s') \, 
D^{-1 *}_{s' s}(R_{\scriptscriptstyle W}(\La, p))
={}&
\sum_{{\scriptscriptstyle{D}'}=1}^4
D^{(1/2, 0) \oplus (0, 1/2) -1}
_{\scriptscriptstyle{D}, \scriptscriptstyle{D}'}
(\La)
\, v_{\scriptscriptstyle D'}(\La p, s) .
\label{By equating the coefficients}
\end{split}
\end{equation}
We now multiply these equations
by the matrices 
$D_{s s''}(R_{\scriptscriptstyle W}(\La, p))$
and
$D^*_{s s''}(R_{\scriptscriptstyle W}(\La, p))$
and sum over $s$
\begin{equation}
\begin{split}
\sum_{s, s'} 
\sqrt{\frac{p^0}{(\La p)^0}} 
u_{\scriptscriptstyle D}(p, s') \, &
D^{-1}_{s' s}(R_{\scriptscriptstyle W}(\La, p))
D_{s s''}(R_{\scriptscriptstyle W}(\La, p))
\\
={}&
\sum_{\scriptscriptstyle{s, D'}}
D^{(1/2, 0) \oplus (0, 1/2) -1}
_{\scriptscriptstyle{D}, \scriptscriptstyle{D}'}
(\La)
\, u_{\scriptscriptstyle D'}(\La p, s) 
D_{s s''}(R_{\scriptscriptstyle W}(\La, p))
\\
\sum_{s, s'}
\sqrt{\frac{p^0}{(\La p)^0}}
\, v_{\scriptscriptstyle D}(p, s') \, &
D^{-1 *}_{s' s}(R_{\scriptscriptstyle W}(\La, p))
D^*_{s s''}(R_{\scriptscriptstyle W}(\La, p))
\\
={}&
\sum_{s, {\scriptscriptstyle{D}'}}
D^{(1/2, 0) \oplus (0, 1/2) -1}
_{\scriptscriptstyle{D}, \scriptscriptstyle{D}'}
(\La)
\, v_{\scriptscriptstyle D'}(\La p, s) 
D^*_{s s''}(R_{\scriptscriptstyle W}(\La, p)).
\label{We then have the coefficients}
\end{split}
\end{equation}
We then get
\begin{equation}
\begin{split}
\sqrt{\frac{p^0}{(\La p)^0}} 
u_{\scriptscriptstyle D}(p, s'') \, 
={}&
\sum_{\scriptscriptstyle{s, D'}}
D^{(1/2, 0) \oplus (0, 1/2) -1}
_{\scriptscriptstyle{D}, \scriptscriptstyle{D}'}
(\La)
\, u_{\scriptscriptstyle D'}(\La p, s) 
D_{s s''}(R_{\scriptscriptstyle W}(\La, p))
\\
\sqrt{\frac{p^0}{(\La p)^0}}
\, v_{\scriptscriptstyle D}(p, s'') \, 
={}&
\sum_{s, {\scriptscriptstyle{D}'}}
D^{(1/2, 0) \oplus (0, 1/2) -1}
_{\scriptscriptstyle{D}, \scriptscriptstyle{D}'}
(\La)
\, v_{\scriptscriptstyle D'}(\La p, s) 
D^*_{s s''}(R_{\scriptscriptstyle W}(\La, p)) .
\label{We then the coefficients}
\end{split}
\end{equation}
We now multiply equations
(\ref{We then the coefficients}) 
by the matrix
$D^{(1/2, 0) \oplus (0, 1/2)}
_{\scriptscriptstyle{D''}, \scriptscriptstyle{D}}
(\La)$
and sum over $\scriptstyle{D}$
and $\scriptstyle{D'}$
\begin{align}
\sum_{\scriptscriptstyle{D}}
\sqrt{\frac{p^0}{(\La p)^0}} &
D^{(1/2, 0) \oplus (0, 1/2)}
_{\scriptscriptstyle{D''}, \scriptscriptstyle{D}}
(\La)
\, u_{\scriptscriptstyle D}(p, s'') \, 
\nn\\
={}&
\sum_{\scriptscriptstyle{s, D, D'}}
D^{(1/2, 0) \oplus (0, 1/2)}
_{\scriptscriptstyle{D''}, \scriptscriptstyle{D}}
(\La)
D^{(1/2, 0) \oplus (0, 1/2) -1}
_{\scriptscriptstyle{D}, \scriptscriptstyle{D}'}
(\La)
\, u_{\scriptscriptstyle D'}(\La p, s) 
D_{s s''}(R_{\scriptscriptstyle W}(\La, p))
\nn\\
\sum_{\scriptscriptstyle{D}}
\sqrt{\frac{p^0}{(\La p)^0}} &
D^{(1/2, 0) \oplus (0, 1/2)}
_{\scriptscriptstyle{D''}, \scriptscriptstyle{D}}
(\La)
\, v_{\scriptscriptstyle D}(p, s'') \, 
\label{By equating the coefficients}
\\
={}&
\sum_{s, {\scriptscriptstyle{D},\scriptscriptstyle{D}'}}
D^{(1/2, 0) \oplus (0, 1/2)}
_{\scriptscriptstyle{D''}, \scriptscriptstyle{D}}
(\La)
D^{(1/2, 0) \oplus (0, 1/2) -1}
_{\scriptscriptstyle{D}, \scriptscriptstyle{D}'}
(\La)
\, v_{\scriptscriptstyle D'}(\La p, s) 
D^*_{s s''}(R_{\scriptscriptstyle W}(\La, p)) .
\nn
\end{align}
These equations are more simply
\begin{equation}
\begin{split}
\sum_{\scriptscriptstyle{D}}
\sqrt{\frac{p^0}{(\La p)^0}} 
D^{(1/2, 0) \oplus (0, 1/2)}
_{\scriptscriptstyle{D''}, \scriptscriptstyle{D}}
(\La)
\, u_{\scriptscriptstyle D}(p, s'') \, 
={}&
\sum_{\scriptscriptstyle{s}}
u_{\scriptscriptstyle D''}(\La p, s) 
D_{s s''}(R_{\scriptscriptstyle W}(\La, p))
\\
\sum_{\scriptscriptstyle{D}}
\sqrt{\frac{p^0}{(\La p)^0}} 
D^{(1/2, 0) \oplus (0, 1/2)}
_{\scriptscriptstyle{D''}, \scriptscriptstyle{D}}
(\La)
\, v_{\scriptscriptstyle D}(p, s'') \, 
={}&
\sum_{s}
v_{\scriptscriptstyle D''}(\La p, s) 
D^*_{s s''}(R_{\scriptscriptstyle W}(\La, p)).
\label{By equating the coefficients}
\end{split}
\end{equation}
We set $\bos p=0$ and 
$p(0) = (m,\bos 0)$,
drop some primes, and interchange
the right- and left-hand sides
of these equations 
\begin{equation}
\begin{split}
u_{\scriptscriptstyle D}(\La p(0), s) 
={}&
\sum_{\scriptscriptstyle{D'}}
\sqrt{\frac{m}{(\La p_0)^0}} 
D^{(1/2, 0) \oplus (0, 1/2)}
_{\scriptscriptstyle{D}, \scriptscriptstyle{D'}}
(\La)
\, u_{\scriptscriptstyle D'}(p(0), s) 
\\
v_{\scriptscriptstyle D}(\La p(0), s) 
={}&
\sum_{\scriptscriptstyle{D'}}
\sqrt{\frac{m}{(\La p)^0}} 
D^{(1/2, 0) \oplus (0, 1/2)}
_{\scriptscriptstyle{D}, \scriptscriptstyle{D'}}
(\La)
\, v_{\scriptscriptstyle D'}(p(0), s) .
\label{By equating the coefficients}
\end{split}
\end{equation}
Setting
$\La = L(p)$, the standard boost
that takes $p(0)$ to $p$, 
\begin{equation}
\begin{split}
u_{\scriptscriptstyle D}(p, s) 
={}&
\sum_{\scriptscriptstyle{D'}}
\sqrt{\frac{m}{(\La p_0)^0}} 
D^{(1/2, 0) \oplus (0, 1/2)}
_{\scriptscriptstyle{D}, \scriptscriptstyle{D'}}
(\La)
\, u_{\scriptscriptstyle D'}(p(0), s) 
\\
v_{\scriptscriptstyle D}(p, s) 
={}&
\sum_{\scriptscriptstyle{D'}}
\sqrt{\frac{m}{(\La p)^0}} 
D^{(1/2, 0) \oplus (0, 1/2)}
_{\scriptscriptstyle{D}, \scriptscriptstyle{D'}}
(\La)
\, v_{\scriptscriptstyle D'}(p(0), s) 
\label{By equating the coefficients}
\end{split}
\end{equation}
and  
switching back to using 3-momenta
to label spinors,  we find
\begin{equation}
\begin{split}
u_{\scriptscriptstyle D}(\bos p, s) 
={}&
\sum_{\scriptscriptstyle{D'}}
\sqrt{\frac{m}{p^0}} 
D^{(1/2, 0) \oplus (0, 1/2)}
_{\scriptscriptstyle{D}, \scriptscriptstyle{D'}}
(\La)
\, u_{\scriptscriptstyle D'}(0, s) 
\\
v_{\scriptscriptstyle D}(\bos p, s) 
={}&
\sum_{\scriptscriptstyle{D'}}
\sqrt{\frac{m}{p^0}} 
D^{(1/2, 0) \oplus (0, 1/2)}
_{\scriptscriptstyle{D}, \scriptscriptstyle{D'}}
(\La)
\, v_{\scriptscriptstyle D'}(0, s) 
\label{By equating the coefficients}
\end{split}
\end{equation}
which are the desired formulas
(\ref{u,v = D u,v0}) that
express the spinors
at finite momentum
in terms of the spinors
at zero momentum.

\section{Majorana and Dirac Fields
\label{Majorana and Dirac Fields sec}}

We have seen (\ref{2-component Majorana fields}--\ref{Dirac field as 2 Majorana fields})
that 
a Dirac field (\ref{4 component Dirac field}) 
is a complex linear
combination of two Majorana fields
(\ref{A Majorana field}) 
of the same mass
\begin{align}
\psi(x) ={}& \frac{1}{\sqrt{2}}
\big[ \psi_{\scriptscriptstyle{M 1}}(x) 
+ i  \psi_{\scriptscriptstyle{M 2}}(x) \big]
\nn \\
={}&
\sum_s \!\! \int \!\! \frac{d^3p}{(2 \pi)^{3/2}} 
\big\{
u(\bos p, s) e^{ip\cdot x} 
\big[ a_1(\bos p,s) + i a_2(\bos p,s) \big]
+
v(\bos p, s) 
e^{-ip\cdot x} 
\big[ a_1^\dag(\bos p,s) + i a_2^\dag(\bos p,s) \big]
\big\}
\nn\\
={}&
\sum_s \int \frac{d^3p}{(2 \pi)^{3/2}} 
\lt[
u_{\scriptscriptstyle D}(\bos p, s) \, e^{ip\cdot x} \, a(\bos p,s)
+
v_{\scriptscriptstyle D}(\bos p, s) \, e^{-ip\cdot x} 
\, a_c^\dag(\bos p, s) \rt]
\label{two Majorana fields}
\end{align}
and that
the annihilation and creation operators
of the Dirac field are complex linear
combinations of the annihilation and creation 
operators of two Majorana fields
$\psi_{\scriptscriptstyle{M 1}}$
and
$\psi_{\scriptscriptstyle{M 2}}$
\begin{equation}
\begin{split}
a(\bos p,s) ={}& \frac{1}{\sqrt{2}}
\Big[ a_1(\bos p,s) + i a_2(\bos p,s) \Big],
\quad
a_c(\bos p,s) ={} \frac{1}{\sqrt{2}}
\Big[ a_1(\bos p,s) - i a_2(\bos p,s) \Big]
\\
a^\dag(\bos p,s) ={}& \frac{1}{\sqrt{2}}
\Big[ a_1^\dag\bos p,s) - i a_2^\dag(\bos p,s) 
\Big],
\quad
a_c^\dag(\bos p,s) ={} \frac{1}{\sqrt{2}}
\Big[ a_1^\dag(\bos p,s) + i a_2^\dag(\bos p,s) 
\Big] .
\end{split}
\end{equation}
\par
The action of a Dirac field is
the sum of the actions of its
Majorana fields
\begin{equation}
{} - \ovl \psi (\c^a \p_a + m ) \psi
=
{} - \thalf \ovl \psi_{\scriptscriptstyle{M 1}} (\c^a \p_a + m ) \psi_{\scriptscriptstyle{M 1}}
{} - \thalf  \ovl \psi_{\scriptscriptstyle{M 2}}  (\c^a \p_a + m ) \psi_{\scriptscriptstyle{M 2}}
\label{sum of the actions} 
\end{equation}
in which $\ovl \psi ={} i \psi^\dag \c^0
= \psi^\dag \b$ and $a = 0, 1, 2, 3$.
The cross-terms
\begin{equation}
- i \thalf \ovl \psi_{\scriptscriptstyle{M 1} }
(\c^a \p_a + m ) \psi_{\scriptscriptstyle{M 2}}
{} + i \thalf  \ovl \psi_{\scriptscriptstyle{M 2}}
(\c^a \p_a + m ) \psi_{\scriptscriptstyle{M 1}}
\label{cross-terms}
\end{equation}
vanish if we integrate by parts
and drop surface terms 
because the fields anticommute
\begin{equation}
\{ \psi^\dag_{\scriptscriptstyle{M 1}} (x), 
\psi_{\scriptscriptstyle{M 2}} (y) \} 
={} 0
\qand
\{ \psi^\dag_{\scriptscriptstyle{M 2}} (x), 
\psi_{\scriptscriptstyle{M 1}} (y) \} 
={} 0 ,
\label{the fields anticommute}
\end{equation}
because they obey the Majorana condition
(\ref{obeys the Majorana condition}),
and because the matrix
$\c^2 \c^0$ is antisymmetric,
while the matrices
$\c^2 \c^0 \c^a$
for $a = 0, 1, 2, 3$ are symmetric
\begin{equation}
(\c^2 \c^0)_{\a \b} 
= {} - (\c^2 \c^0)_{\b \a}
\qand
(\c^2 \c^0 \c^a)_{\a \b} 
= {}  (\c^2 \c^0 \c^a)_{\b \a}.
\label{antisymmetric/symmetric}
\end{equation}
\par
By using the formulas
(\ref{the gamma matrices})
for the gamma matrices,
we may write the action 
for a single Majorana field as
\begin{equation}
\begin{split}
- \thalf \ovl \psi_{\scriptscriptstyle{M}} 
(\c^a \p_a + m ) \psi_{\scriptscriptstyle{M}}
={}&
- \thalf i \psi_{\scriptscriptstyle{M}}^\dag \c^0
(\c^a \p_a + m ) \psi_{\scriptscriptstyle{M}}
\\
={}&
    - \thalf i \psi_{\scriptscriptstyle{M}}^\dag 
(- \p_0 + \c^0 \bos \c \cdot \bos \grad 
    + m \c^0 ) \psi_{\scriptscriptstyle{M}} .
\label{a Majorana action}
\end{split}
\end{equation} 
Like a Dirac field,
a 4-component Majorana field is composed of 
two 2-component fields~\citep{Dreiner:2008tw}
\begin{equation}
\psi_{\scriptscriptstyle{M}} = {} 
\begin{pmatrix}
\ell \\ r 
\end{pmatrix},
\label{two 2-component fields}
\end{equation}
in which $\ell$ is left handed
and $r$ is right handed.
In terms of $\ell$ and $r$, 
the action (\ref{a Majorana action}) 
of the Majorana field may be written 
as~\citep{Dreiner:2008tw,CahillCUP2DiracField}
\begin{equation}
\begin{split}
- \thalf \ovl \psi_{\scriptscriptstyle{M}} 
(\c^a \p_a + m ) \psi_{\scriptscriptstyle{M}}
={}&
\thalf i \begin{pmatrix}
\ell^\dag & r^\dag
\end{pmatrix}
\begin{pmatrix}
\p_0 - \bos \s \cdot \bos \grad & i m \\
i m & \p_0 + \bos \s \cdot \bos \grad \\
\end{pmatrix} 
\begin{pmatrix}
\ell \\
r \\
\end{pmatrix}
\\
={}&
\thalf i \ell^\dag 
(\p_0 - \bos \s \cdot \bos \grad) \ell
+
\thalf i r^\dag 
(\p_0 + \bos \s \cdot \bos \grad) r
- \thalf m (\ell^\dag r + r^\dag \ell).
\label{the action of a Majorana field}
\end{split}
\end{equation}
In this notation, 
the Majorana condition
(\ref{obeys the Majorana condition})
\begin{equation}
\begin{pmatrix}
\ell^* \\ r^* \\
\end{pmatrix}
= - i \begin{pmatrix}
0 & \s^2 \\
- \s^2 & 0 \\
\end{pmatrix}
\begin{pmatrix}
\ell \\ r \\
\end{pmatrix}
=
\begin{pmatrix}
- i \s^2 r \\
i \s^2 \ell \\
\end{pmatrix}
\end{equation}
tells us that $\ell ={} - i \s^2 r^*$
and $r ={} i \s^2 \ell^*$,
or more simply that
$\ell_1 = - r_2^*$ and $\ell_2 = r_1^*$.
So one can write the action
(\ref{the action of a Majorana field})
of a Majorana field entirely
in terms of $\ell$
\begin{equation}
\begin{split}
- \thalf \ovl \psi_{\scriptscriptstyle{M}} 
(\c^a \p_a + m ) \psi_{\scriptscriptstyle{M}}
={}&
\thalf i \ell^\dag 
(\p_0 - \bos \s \cdot \bos \grad) \ell
+
\thalf i \ell\transpose 
(\p_0 - \bos \s^* \cdot \bos \grad)  \ell^*
- \thalf i m 
(\ell^\dag \s^2 \ell^* - \ell\transpose \s^2 \ell)
\end{split}
\end{equation}
or entirely in terms of $r$
\begin{align}
- \thalf \ovl \psi_{\scriptscriptstyle{M}} 
(\c^a \p_a + m ) \psi_{\scriptscriptstyle{M}}
={}
\thalf i r\transpose 
(\p_0 + \bos \s^* \cdot \bos \grad) r^*
+
\thalf i r^\dag 
(\p_0 + \bos \s \cdot \bos \grad) r
- 
\thalf i m 
(r\transpose \s^2 r - r^\dag \s^2 r^*) .
\end{align}
The action (\ref{the action of a Majorana field})
takes simpler forms when we
integrate by parts, anticommute
the fields, and drop both the surface terms
and an infinite constant
\begin{equation}
\begin{split}
- \thalf \ovl \psi_{\scriptscriptstyle{M}} 
(\c^a \p_a + m ) \psi_{\scriptscriptstyle{M}}
={}&
i \ell^\dag 
(\p_0 - \bos \s \cdot \bos \grad) \ell
- \thalf i m 
(\ell^\dag \s^2 \ell^* - \ell\transpose \s^2 \ell)
\\
={}&
i r^\dag 
(\p_0 + \bos \s \cdot \bos \grad) r
- 
\thalf i m 
(r\transpose \s^2 r - r^\dag \s^2 r^*) .
\end{split}
\end{equation}
\par
Under a Lorentz transformation $L$,
the fields $\ell(x)$ and $r(x)$
transform as
\begin{equation}
\begin{split}
U(L) \, \ell(x) \, U^{-1}(L) ={}& 
D^{(1/2,0)} (L^{-1})
\, \ell(Lx)
\\
U(L) \, r(x) \, U^{-1}(L) ={}& 
D^{(0,1/2)} (L^{-1})
\, r(Lx)
\label {left-and-right-handed Weyl spinors}
\end{split}
\end{equation}
in which the unitary operator
$ U(L) \equiv U(L(\bos \th, \bos \l)) $
and the complex, unimodular 
$2 \by 2$ matrices
of unit determinant
\begin{equation}
D^{(1/2,0)}(L(\bos{\theta},\bos{\lambda }) )
= e^{\mbox{} - (\bos \l + i \bos \th) \cdot \bos{\s}/2} 
\qand
D^{(0,1/2)}(L(\bos{\thet},\bos{\lambda }) )
= e^{\mbox{}  (\bos \l - i \bos \th) \cdot \bos{\s}/2}
\label {D1/20thet lam 2}
\end{equation} 
both represent the 
Lorentz transformation
\begin{equation}
L(\bos \th, \bos \l) ={}
e^{\bos \thet \cdot \bos R 
+ \bos \lambda \cdot \bos B}
\end{equation}
where
\begin{equation}
R_1 = \bpm 
           0 & 0 & 0 & 0 \\
           0 & 0 & 0 & 0 \\
           0 & 0 & 0 & -1 \\
           0 & 0 & 1 & 0 \\
           \epm \quad
R_2 = \bpm 
           0 & 0 & 0 & 0 \\
           0 & 0 & 0 & 1 \\
           0 & 0 & 0 & 0 \\
           0 & -1 & 0 & 0 \\
           \epm \quad
R_3 = \bpm 
           0 & 0 & 0 & 0 \\
           0 & 0 & -1 & 0 \\
           0 & 1 & 0 & 0 \\
           0 & 0 & 0 & 0 \\
           \epm
\label {3Rs}
\end{equation}
and
\begin{equation}
B_1 = \bpm 
           0 & 1 & 0 & 0 \\
           1 & 0 & 0 & 0 \\
           0 & 0 & 0 & 0 \\
           0 & 0 & 0 & 0 \\
           \epm \quad
B_2 = \bpm 
           0 & 0 & 1 & 0 \\
           0 & 0 & 0 & 0 \\
           1 & 0 & 0 & 0 \\
           0 & 0 & 0 & 0 \\
           \epm \quad
B_3 = \bpm 
           0 & 0 & 0 & 1 \\
           0 & 0 & 0 & 0 \\
           0 & 0 & 0 & 0 \\
           1 & 0 & 0 & 0 \\
           \epm .
\label {3Bs}
\end{equation}
For small $ \bos \th $ and $\bos \l$,
the Lorentz transformation
$L(\bos \th, \bos \l)$
changes $t, \bos x$ to
\bea
t' & \simeq & t + 
\bos{\lambda} \cdot \bos{x} \nn\\
\bos{x}' & \simeq & \bos{x} + t \bos{\lambda}
+ \bos{\thet \wedge x}
\label {x' = Lx succinctly}
\eea
in which $\wedge \equiv \times$ 
means cross-product.

\bibliography{physics,math,lattice,books}
\end{document}